\documentclass[%
reprint,
superscriptaddress,
showkeys,
amsmath,amssymb,
aps,
]{revtex4-2}

\usepackage{graphicx}
\usepackage{dcolumn}
\usepackage{bm}
\usepackage{hyperref}
\usepackage{verbatim}
\usepackage[utf8]{inputenc}
\usepackage{layouts}
\usepackage{multirow}
\usepackage[dvipsnames]{xcolor}
\usepackage{hhline} 
\usepackage[separate-uncertainty=true]{siunitx}
\usepackage[version=3]{mhchem} 
\usepackage{wrapfig}
\usepackage{braket}
\usepackage{float}
\usepackage{leftidx} 
\usepackage{amsmath,mathtools}
\usepackage{amssymb}
\usepackage{chngcntr}

\hypersetup{colorlinks=true,urlcolor=magenta,linkcolor=magenta,citecolor=cyan}
\begin{document}

\title{Breakdown of the static dielectric screening approximation of Coulomb interactions in atomically thin semiconductors}

\author{Amine Ben Mhenni}
\email{Amine.Ben-Mhenni@tum.de}
\affiliation{%
    Walter Schottky Institute, TUM School of Natural Sciences, and MCQST, Technical University of Munich, Munich, Germany
}%

\author{Dinh Van Tuan}%

\affiliation{%
    Department of Electrical and Computer Engineering, University of Rochester, Rochester, NY, United States.
}%

\author{Leonard Geilen}%
\affiliation{%
    Walter Schottky Institute, TUM School of Natural Sciences, and MCQST, Technical University of Munich, Munich, Germany
}%

\author{Marko M. Petrić}%
\affiliation{%
    Walter Schottky Institute, TUM School of Computation, Information and Technology, and MCQST, Technical University of Munich, Munich, Germany
}%

\author{Melike Erdi}%
\affiliation{%
    School for Engineering of Matter, Transport and Energy, Arizona State University, Tempe, AZ, United States.
}%

\author{Kenji Watanabe}%
\affiliation{%
 Research Center for Electronic and Optical Materials, National Institute for Materials Science, 1-1 Namiki, Tsukuba 305-0044, Japan
}%

\author{Takashi Taniguchi}%
\affiliation{%
 Research Center for Materials Nanoarchitectonics, National Institute for Materials Science,  1-1 Namiki, Tsukuba 305-0044, Japan
}%

\author{Sefaattin Tongay}%
\affiliation{%
 School for Engineering of Matter, Transport and Energy, Arizona State University, Tempe, AZ, United States.
}%

\author{Kai Müller}%
\affiliation{%
    Walter Schottky Institute, TUM School of Computation, Information and Technology, and MCQST, Technical University of Munich, Munich, Germany
}%

\author{Nathan P. Wilson}%
\affiliation{%
    Walter Schottky Institute, TUM School of Natural Sciences, and MCQST, Technical University of Munich, Munich, Germany
}%

\author{Jonathan J. Finley}%
\email{JJ.Finley@tum.de}
\affiliation{%
    Walter Schottky Institute, TUM School of Natural Sciences, and MCQST, Technical University of Munich, Munich, Germany
}%

\author{Hanan Dery}%
\affiliation{%
    Department of Electrical and Computer Engineering, University of Rochester, Rochester, NY, United States.
}%
\affiliation{%
    Department of Physics and Astronomy, University of Rochester, Rochester, NY, United States.
}%

\author{Matteo Barbone}%
\email{Matteo.Barbone@wsi.tum.de}
\affiliation{%
    Walter Schottky Institute, TUM School of Computation, Information and Technology, and MCQST, Technical University of Munich, Munich, Germany
}%

\date{\today}

\begin{abstract}
Coulomb interactions in atomically thin materials are uniquely sensitive to variations in the dielectric screening of the environment, which can be used to control exotic quantum many-body phases and engineer exciton potential landscapes. A static approximation of the dielectric response, where increased dielectric screening is predicted to cause an energy redshift of the exciton resonance, has been until now sufficient. Here, we use charge-tunable exciton resonances to study screening effects in transition metal dichalcogenide monolayers embedded in materials with dielectric constants ranging from $4$ to more than $1000$, a range two orders of magnitude larger than previous studies. In contrast to the redshift predicted by static models employed until now, we observe a blueshift of the exciton resonance exceeding $30$ meV for larger dielectric constant environments. By introducing a dynamical screening model based on a new solution to the Bethe-Salpeter equation, we find that while the exciton binding energy remains mostly controlled by the static dielectric response, the exciton self-energy is dominated by the high-frequency one. Our results supplant the understanding of screening in layered materials and their heterostructures, introduce a knob to tune selected many-body effects, and open new routes to detect and control correlated quantum many-body states and to design optoelectronic and quantum devices.
\end{abstract}

\maketitle


\subsection*{Introduction}

Interactions amongst particles give rise to collective phenomena described by new fundamental laws beyond simplified single-particle systems \cite{snoke_solid_2019}. This is particularly evident in heterostructures of two-dimensional ($2$D) materials, in which a wide variety of correlated electronic and excitonic phases have been realized, driven by strong Coulomb interactions \cite{yankowitz_van_2019, mak_semiconductor_2022, wilson_excitons_2021, regan_emerging_2022, montblanch_layered_2023}. For instance, excitonic complexes up to eight particles \cite{barbone_charge-tuneable_2018, sidler_fermi_2017, van_tuan_six-body_2022} and signatures of Wigner crystals \cite{smolenski_signatures_2021} have recently been reported in encapsulated, gated monolayer transition metal dichalcogenides (TMDs). Hubbard physics \cite{tang_simulation_2020,xu_correlated_2020}, unconventional superconductivity \cite{cao_unconventional_2018}, and Chern insulators \cite{chen_tunable_2020} have been observed in moiré superlattices. 

In all such phenomena, Coulomb interactions are heavily influenced by the dielectric response of the environment because the electric field created by charge quasiparticles in a 2D material extends into the surrounding medium \cite{rytova_screened_2020, keldysh_coulomb_1979, cudazzo_dielectric_2011, chernikov_exciton_2014, ugeda_giant_2014}. 
This in turn leads to large exciton binding energy and bandgap renormalisation effects\cite{ugeda_giant_2014}, with Coulomb engineering of atomically thin monolayers attracting considerable interest as a deterministic, scalable, and clean route to control many-body states, from exciton localisation and transport to tuning many-body interactions in correlated states \cite{raja_coulomb_2017, forsythe_band_2018}. To describe screening in semiconductor quantum wells and 2D materials, a common practice is to use an effective dielectric constant, neglecting frequency dependence and greatly simplifying the description of interactions between quasiparticles \cite{van_tuan_coulomb_2018, cho_environmentally_2018}. Within the limit of small variations (below an order of magnitude) in the dielectric constants of the environments studied so far, dynamical screening effects did not appear necessary to describe excitons in TMD monolayers \cite{chernikov_exciton_2014, raja_coulomb_2017, raja_dielectric_2019, stier_probing_2016, tebbe_tailoring_2023, goryca_revealing_2019, raja_dielectric_2019}. 

Here, we track gate-tunable exciton resonances in monolayer WSe$_{2}$ embedded in environments with static dielectric constants spanning three orders of magnitude but with the high-frequency dielectric constant changing by less than two times. In contrast with the preceding literature, we surprisingly observe an exciton resonance \textit{blueshift} for larger static dielectric constants, incompatible with the established theoretical understanding. We explain this behavior by introducing a model that accounts for the dynamic screening of electron-hole bound states, which shows that the exciton binding energy primarily responds to the static dielectric constant, while the self-energy of the bound state primarily depends on the high-frequency dielectric constant. Crucially, the free-particle bandgap remains dependent on the static dielectric constant, and reveals its inadequacy to describe bound electron-hole pairs under more extreme screening conditions. Our results reveal conditions under which the frequency-independent dielectric screening approximation breaks down, and dynamic effects become a key factor in determining excitonic behaviour. Further, they indicate the necessity of addressing excitons as bound-states to fully capture screening in $2$D systems. Materials with strong frequency-dependent dielectric functions allow the selective tuning of exciton binding energy and self-energy, providing a knob to control quantum many-body states and their interactions and to design dielectric engineered optoelectronic and quantum devices.

\subsection*{\label{sec:optical_probing}Effect of the dielectric screening on the optical spectrum of monolayer TMDs}

\begin{figure*}
\includegraphics[width=1\textwidth]{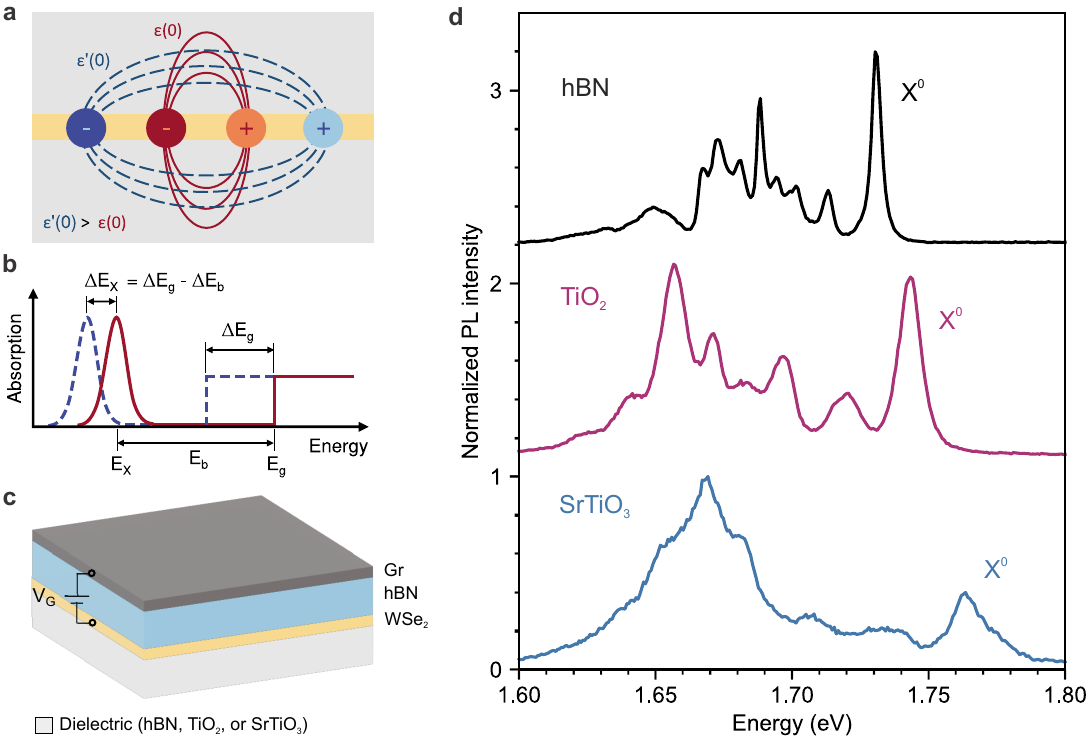}
\caption{\label{fig:config}\textbf{Dielectric screening in a monolayer semiconductor and device configurations.}
\newline
\textbf{a}, Schematic of an exciton and the electric field lines between its electron and hole when an atomically thin semiconductor is embedded in a weak (strong) screening environment whose effective dielectric constant is $\varepsilon (0)$ ($\varepsilon' (0)$). \textbf{b}, Sketches of the expected absorption spectra where $\mathrm{E_{X}}$ denotes the energy resonance of the exciton ground state (n $= 1$), $\mathrm{E_{b}}$ the binding energy, and $\mathrm{E_{g}}$ the continuum bandgap energy (exciton energy in the limit n $= \infty$). \textbf{c}, Schematic of our gate-tunable devices for the study of Coulomb interactions. In each device, a monolayer WSe$_{2}$ is placed between hBN and a bottom dielectric, which is either hBN, TiO$_{2}$, or SrTiO$_{3}$. \textbf{d}, Normalized PL spectra of the different dielectric configurations at $8$ K. The resonance energy of the charge-neutral exciton X$^{0}$ blueshifts with increasing $\varepsilon (0)$ of the environment.}
\end{figure*}

Figure~\ref{fig:config}a shows the schematic of an exciton in an atomically thin semiconductor embedded in environments with two different effective static dielectric constants $\varepsilon (\omega = 0)$ and $\varepsilon' (0)$, where $\varepsilon (0)<\varepsilon' (0)$. Exciton states manifest as discrete optical resonances below the renormalized free-particle bandgap energy as shown in Fig.~\ref{fig:config}b for the exciton ground state. The dielectric environment affects the exciton resonance energy through changes to both its binding energy and its self-energy (the energy accounting for all interactions), the latter being a bandgap renormalization (BGR) effect of the bound electron-hole pair. For increasing effective dielectric constants, the quasiparticle self-energy reduces, inducing a \textit{redshift} of the exciton resonance. At the same time, the exciton binding energy also decreases, thereby inducing a \textit{blueshift} of the exciton resonance. Scanning tunneling spectroscopy experiments, which measure the free-particle bandgap, and optical absorption, revealed the two effects to be of the same order of magnitude in TMDs (up to $\sim$hundreds of meV), almost cancelling each other \cite{ugeda_giant_2014, raja_coulomb_2017}. However, in the static approximation, the former is expected to be always slightly stronger than the latter by up to a few tens of meV \cite{scharf_dynamical_2019, cho_environmentally_2018}. In fact, when calculating the BGR, the Coulomb potential $\Delta V(r)$ is evaluated at a distance $r\rightarrow 0$, whereas the binding energy is evaluated at a finite distance. Since the difference between the Coulomb potentials in two dielectric environments is greatest at $r = 0$, the net effect should always be a redshift of the exciton resonance with increasing static dielectric constant \cite{scharf_dynamical_2019, cho_environmentally_2018}. Importantly, this picture also implies that static screening alone does not allow to independently tune binding energy and self-energy. To date, applications using dielectric engineering to control quasiparticles and their interactions, as well as to design devices, have rested on this understanding.

We fabricate charge-tunable devices based on monolayer WSe$_{2}$ by using van der Waals fabrication techniques (Methods). In this study, we use WSe$_{2}$ as a prototypical TMD material since it offers a larger exciton Bohr radius than Mo-based TMDs \cite{stier_magnetooptics_2018,goryca_revealing_2019}, amplifying its sensitivity to the dielectric environment and because it does not display significant Fermi level pinning \cite{barbone_charge-tuneable_2018, he_valley_2020}. Figure \ref{fig:config}c shows the device configuration. Monolayer WSe$_{2}$ is sandwiched between a top layer of hexagonal boron nitride (hBN) and a bottom dielectric with varying $\varepsilon (0)$, either hBN, TiO$_{2}$, or SrTiO$_{3}$. Throughout this work, we will refer to the different dielectric configurations by their bottom dielectric layer. At temperatures $\leq 10$ K, the effective $\varepsilon (0)$ of these configurations range from  $\sim3.5$ for the hBN \cite{smolenski_signatures_2021} sample, to $\sim 75$ for the TiO$_{2}$ sample \cite{schoche_infrared_2013}, and $>1000$ for the SrTiO$_{3}$ sample \cite{neville_permittivity_1972}, spanning a range more than two orders of magnitude higher than previous studies \cite{chernikov_exciton_2014, raja_coulomb_2017, raja_dielectric_2019, stier_probing_2016, tebbe_tailoring_2023, goryca_revealing_2019,raja_dielectric_2019}. Few-layer graphene (Gr) is used as a gate, allowing tuning of the electrochemical potential in all devices. Figure \ref{fig:config}d shows the low-temperature photoluminescence (PL) spectra near charge neutrality, evidenced by the high ratio between neutral exciton X$^{0}$ and negative trion X$^{-}$ intensities. In contrast to previous reports \cite{stier_probing_2016, raja_coulomb_2017, tebbe_tailoring_2023}, the X$^{0}$ energy surprisingly blueshifts with increasing effective dielectric constant, from $1.731$ eV in the hBN sample, to $1.743$ eV in the TiO$_{2}$ sample, and further to $1.764$ eV in the SrTiO$_{3}$ sample. These findings are not limited to selected WSe$_{2}$ samples, but we observe consistent blueshifts across more than $12$ samples embedded in the same dielectric environments, including in monolayer MoSe$_{2}$ and WS$_{2}$ crystals (Supplementary Fig. $1$). 

\begin{figure*}
\includegraphics[width=1\textwidth]{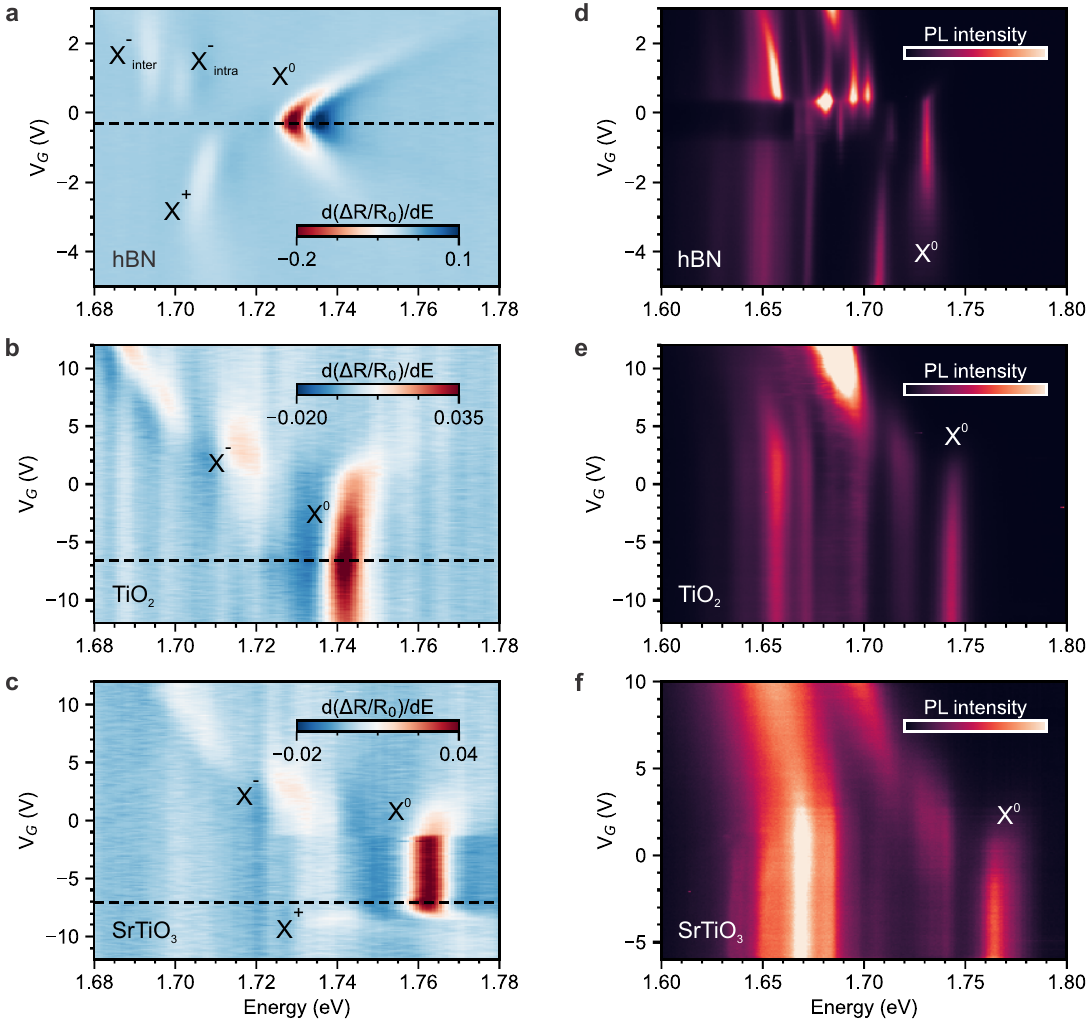}
\caption{\label{fig:gated}\textbf{Gate-dependent optical response in different screening environments.}
\newline
\textbf{a-c}, Gate-dependent reflection contrast derivative ($d(\Delta R / R_{0}) / dE)$) of monolayer WSe$_{2}$ in the hBN (\textbf{a}), TiO$_{2}$ (\textbf{b}), and SrTiO$_{3}$ (\textbf{c}) dielectric configuration. The voltage corresponding to charge neutrality is indicated by a dashed horizontal line. X$^{0}$ and the charged excitons are labeled in the figures. At charge neutrality, X$^{0}$ blueshifts with increasing effective static dielectric constant of the environment. \textbf{d-f}, Gate-dependent PL spectra of monolayer WSe$_{2}$ in the hBN (\textbf{d}), TiO$_{2}$ (\textbf{e}), and SrTiO$_{3}$ (\textbf{f}) dielectric configuration. X$^{0}$ is labeled in the figures.}
\end{figure*}

In contrast with past studies, we measure the gate-dependent optical response of monolayer WSe$_{2}$ in the different dielectric configurations to exclude possible contributions to the exciton resonance shift from charge doping \cite{van_tuan_probing_2019}. Figures \ref{fig:gated}a-c compare the gate-dependent reflection contrast derivatives ($d(\Delta R / R_{0}) / dE)$) from the hBN, TiO$_{2}$, and SrTiO$_{3}$ samples. In all cases, we extract the X$^{0}$ energy by fitting a dispersive Lorentzian at the charge neutrality point identified from the X$^{0}$ absorption maximum (Supplementary Fig. $2$). In the hBN sample (Fig. \ref{fig:gated}a), the energy of X$^{0}$ is $1.731$ eV.
The spectrum of X$^{0}$ exhibits a pronounced broadening and energy blueshift larger than 15 meV from charge neutrality to higher charge doping. This highlights the importance of evaluating excitonic energies at charge neutrality in such studies. The negative exchange-split trions \cite{barbone_charge-tuneable_2018} (X$^{-}_{\mathrm{intra}}$ and X$^{-}_{\mathrm{inter}}$) appear in the electron-doped regime (positive $V_{\mathrm{G}}$). In contrast, the positively charged trion \cite{barbone_charge-tuneable_2018} (X$^{+}$) becomes visible in the hole-doped regime (negative $V_{\mathrm{G}}$). The TiO$_{2}$ sample (Fig. \ref{fig:gated}b) shows the X$^{0}$ at $1.740$ eV, $9$ meV blueshifted with respect to X$^{0}$ in the hBN sample. Even more, the SrTiO$_{3}$ sample (Fig. \ref{fig:gated}c) shows a X$^{0}$ energy of $1.762$ eV, $31$ meV blueshifted with respect to that in the hBN sample.

To further corroborate our findings, we inspect the optical response of WSe$_{2}$ via gate-dependent PL spectroscopy, and extract the energy of X$^{0}$ at charge neutrality. Figures \ref{fig:gated}d-f show the gate-dependent PL spectra of the hBN, TiO$_{2}$, and SrTiO$_{3}$ samples. In the hBN sample (Fig. \ref{fig:gated}d), the energy of X$^{0}$ is $1.731$ eV, accompanied by a linewidth below $2$ meV, consistent with the highest quality samples reported in the literature \cite{barbone_charge-tuneable_2018, he_valley_2020, liu_magnetophotoluminescence_2019, liu_exciton-polaron_2021}, and blueshifts due to charge doping by up to $5$ meV before disappearing (Supplementary Fig. $3$). The excited states $2s$, $3s$, $4s$, and $(2s)^{+}$ are well-resolved in the PL spectra (Supplementary Fig. $4$),  further testifying to the high sample quality \cite{liu_magnetophotoluminescence_2019,liu_exciton-polaron_2021}. In the TiO$_{2}$ sample (Fig. \ref{fig:gated}e), X$^{0}$ has a linewidth of less than $5$ meV and appears at $1.743$ eV, blueshifted by $\sim12$ meV compared to that in the hBN sample. In the SrTiO$_{3}$ sample (Fig. \ref{fig:gated}f), X$^{0}$ has a linewidth of $\sim6$ meV and arises at $1.764$ eV, blueshifted by $\sim33$ meV compared to that in the hBN sample. Overall, PL measurements are in good agreement with the reflection contrast data.

To exclude any contribution to the exciton energy shifts from uncontrolled strain fields \cite{aslan_strain_2018} or other spatially dependent effects, we study the X$^{0}$ energy distribution over large areas on multiple samples for each dielectric configuration (Supplementary Fig. $5$). We observe a narrow distribution below $3$ meV, reflecting the high homogeneity of the samples and the repeatability of the observations.

Since $\varepsilon\mathrm{^{SrTiO_{3}}} (0)$ increases over one order of magnitude between 100 K and 5 K \cite{neville_permittivity_1972}, we also look at the temperature dependence of the exciton resonance in the SrTiO$_3$ device (Supplementary Fig. $6$). 
Going from $80$ K down to $20$ K, X$^{0}$ exhibits a blue shift ($\sim 23$ meV) which is more than twice larger than the blue shift in the hBN sample ($\sim 9$ meV).
This presents further evidence that X$^{0}$ blueshifts with increasing $\varepsilon (0)$ of the environment.
Moreover, it unveils a new pathway to control X$^{0}$ on the same device by tuning the $\varepsilon (0)$ of SrTiO$_{3}$ via temperature or via electric fields \cite{neville_permittivity_1972}.

\subsection*{\label{sec:theory}Fully dynamical description of Coulomb screening}

\begin{figure*}
\includegraphics[width=1\textwidth]{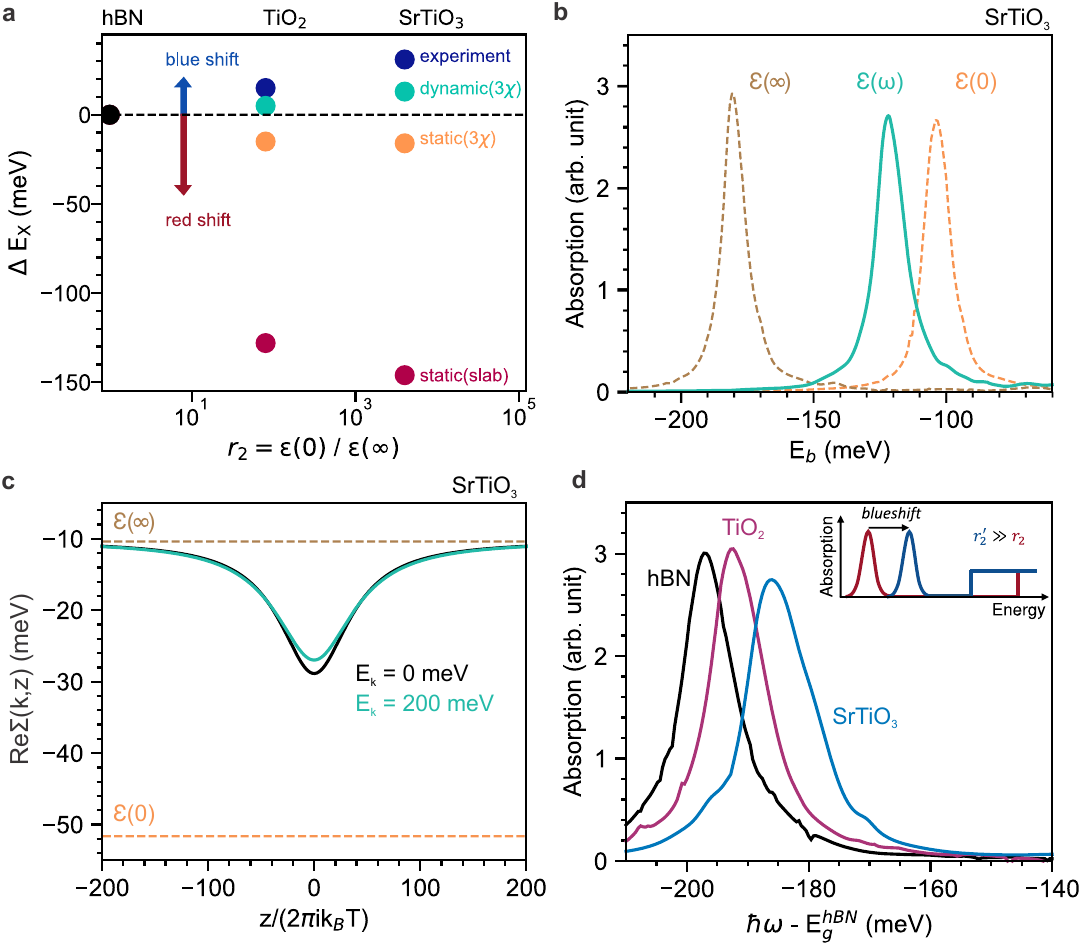}
\caption{\label{fig:theory}\textbf{Modelling dynamical dielectric screening effects.}
\newline
\textbf{a}, Comparison of the energy shift of X$^{0}$ in monolayer WSe$_2$ as a function of the parameter $r_{2}$ from the optical experiments and from the theoretical calculations for both static and dynamical models. Static models always lead to redshift, with the slab model diverging from the experimental results by almost 180 meV. \textbf{b}, Exciton binding energy calculated from the absorption spectrum of monolayer WSe$_{2}$ on SrTiO$_{3}$ by neglecting the self-energy term in the BSE equation and by using $\varepsilon (0)$ (red), $\varepsilon (\infty)$ (blue), and $\varepsilon (\omega)$ (green). \textbf{c}, Calculated real part of the self-energy $\Sigma (k,z)$ of conduction band electrons as a function of Matsubara frequencies for the SrTiO$_{3}$ sample. Solid lines indicate dynamical calculations for two relevant electron energies, while dashed lines indicate static calculations of the single-particle BGR calculated with $\varepsilon (\infty)$ and and $\varepsilon (0)$.Single-particle BGR calculated with $\varepsilon (\omega)$ (not shown) and $\varepsilon (0)$ lead to almost identical results. Zero energy is set to the calculated self-energy of the hBN sample for $\varepsilon (\infty)$. \textbf{d}, Absorption spectra corresponding to the samples measured experimentally calculated with dynamical screening ($\varepsilon (\omega)$) by including both binding energy and self-energy of the bound states. The spectra are plotted relative to $\mathrm{E_{g}^{hBN}} = 1.9$ eV. In the inset, schematic of the absorption spectrum of a monolayer TMD highlighting the exciton resonance blueshift when $r_{2}$ increases from $\sim1$ to $\gg1$.}
\end{figure*}

We compare our experimental results with the theoretical predictions from two models employed to describe the influence of the environment on exciton resonances in TMDs in the static screening approximation, the "slab" model \cite{cho_environmentally_2018} and the "3$\chi$" model \cite{van_tuan_coulomb_2018}, and track the predicted exciton resonance shift with varying screening $r_{2} = \varepsilon (0)/\varepsilon (\infty)$ from the reference point of an hBN environment. Figure \ref{fig:theory}a shows that with increasing $r_{2}$, exciton resonances according to the slab and 3$\chi$ model are expected to redshift from the hBN reference up to about 16 meV and 145 meV respectively, or about 45-170 meV lower in energy than our experimental results. The large difference between the two models stems from the lower screening weight attributed to the surrounding environment by the 3$\chi$. For a homogeneous strain field to be the source of such a shift, that would amount to a compressive strain $\sim 1-4\%$ \cite{schmidt_reversible_2016}, which has never been reported even for externally applied deformation, while the adhesion energy of WSe$_2$ to the substrate would only support a planar strain well below 0.1$\%$ before delamination\cite{blundo_experimental_2021}. 
Also, the energy of the ground state (1s) exciton resonance is less sensitive to strain than other established experimental routes, such as the relative energy between the 1s and the 2s exciton \cite{aslan_strain_2018}, which is employed as a more direct measurement of the binding energy and the electronic bandgap \cite{raja_coulomb_2017}.
Having excluded other potential sources of blueshift, we conclude that X$^{0}$ blueshifts with an increasing static dielectric constant of the environment. This implies that the corresponding reduction in the exciton binding energy must be greater than the BGR. Hence, the static approximation of Coulomb interactions is not sufficient to describe dielectric screening in atomically thin semiconductors.

To reconcile the contradiction between our results and the theory of screened many-body interactions, we turn to examining the role of frequency dependence in dielectric screening. The response of a dielectric material to an electric field comes from its valence electrons and, if the material is polar, from field-induced lattice vibrations that induce a net polarization \cite{lyddane_polar_1941}. The electron and hole are not independent entities; instead, they move with respect to each other with kinetic energy commensurate with the exciton binding energy as dictated by the virial theorem, resulting in a rapidly varying electric field \cite{van_tuan_coulomb_2018, scharf_dynamical_2019}. Consequently, we lift the assumption that the atoms of the encapsulating layers either perfectly trace or completely ignore the fast variation of the electric field. If the dielectric layer adjacent to the monolayer semiconductor is a polar material, we can approximate its response to electric field at frequency $\omega$ by the dielectric function:

\begin{eqnarray*}
  \epsilon(\omega) = \epsilon (\infty) \,\,\prod_j \frac{\omega_{j,\text{LO}}^2 - \omega^2}{\omega_{j,\text{TO}}^2 - \omega^2}\,\, 
  \label{eq:LST}
\end{eqnarray*}

The ratio between the static and high-frequency dielectric constants is the Lyddane–Sachs–Teller relation $r_{2} = \epsilon(0)/\epsilon(\infty) = \prod_j \omega_{j,\text{LO}}^2 /\omega_{j,\text{TO}}^2 $ \cite{lyddane_polar_1941}. The index $j$ runs over the optical phonon modes, where $\omega_{j,\text{LO/TO}}$ is the associated frequency of the longitudinal/transverse optical lattice vibration in the dielectric layers. 
In the following, we use the $3\chi$ formulation of the Coulomb potential \cite{van_tuan_coulomb_2018} and introduce dynamical dielectric functions to model the response of top and bottom dielectrics. We calculate the dynamical self-energy of conduction band electrons from the solution of the Dyson equation \cite{dyson_s_1949} (Supplementary Theoretical Methods).

Figure \ref{fig:theory}a shows the effect of dynamical screening on excitons when $r_{2} \gg 1$. Consistent with our measurements, X$^{0}$ blueshifts because the binding energy blueshift contribution $\mathrm{\Delta E_{b}}$ is larger than the redshift contribution $\mathrm{\Delta E_{g}}$ for a higher $r_{2}$. To understand the physical reasons behind such results, we consider individually each contribution.

Figure \ref{fig:theory}b shows the binding energy of the SrTiO$_{3}$ sample calculated by neglecting the self-energy term from the absorption spectrum. The binding energy calculated with the static dielectric constant $\varepsilon (0)$, which assumes that atoms can readily trace the varying electric field of the electron and hole, is $\sim 80$ meV smaller than that calculated with the high-frequency dielectric constant $\varepsilon (\infty)$, which only considers the electronic contribution to the screening. Calculating the binding energy by using the dynamical dielectric function $\varepsilon (\omega)$ in the effective Bethe-Salpeter equation (BSE) \cite{salpeter_relativistic_1951}, we obtain results closer to $\varepsilon (0)$, indicating that the binding energy is mostly influenced by the static dielectric constant. 

Figure \ref{fig:theory}c shows the real part of the self-energy $\Sigma(k,z)$ for the SrTiO$_{3}$ sample as a function of imaginary Matsubara frequencies. The reference point at $0$ meV is set to the self-energy calculated for the hBN sample with $\varepsilon (\infty)$. The self-energies calculated for the SrTiO$_{3}$ sample at $\varepsilon (0)$ and $\varepsilon (\infty)$ are $\sim 40$ meV apart. We calculate the dynamical self-energy for two relevant electron energies $E=\hbar^2k^2/2m_c$ of $0$ and $200$ meV. Across the whole Matsubara frequency spectrum, the self-energies remain close to the value calculated with $\varepsilon (\infty)$, indicating that the dynamical self-energy is mainly influenced by the high-frequency dielectric constant. Importantly, the value of the single-particle BGR calculated with $\varepsilon (\omega)$ is $\sim$50 meV, almost identical to the BGR calculated with $\varepsilon (0)$ and opposite to the results obtained by considering the self-energy of the bound pair. 

Figure \ref{fig:theory}d presents the calculated absorption spectra of X$^{0}$ by including both self-energy and binding energy in the BSE for all the dielectric configurations considered in our experiments. The results show a net blueshift with increasing dielectric constant, in agreement with the experimental findings.

Despite the qualitative agreement of our theoretical and experimental results, we note a lower calculated shift, possibly due to underestimation of the environmental screening in the $3\chi$ model \cite{van_tuan_coulomb_2018}, as well as a possible smaller difference of the high-frequency dielectric constants (that is $\varepsilon\mathrm{^{SrTiO_{3}}}  (\infty) - \varepsilon\mathrm{^{hBN}} (\infty) < 3.2)$. 
We also underline that we are unable to obtain a blueshift from the slab model even with dynamical screening: the much larger weight attributed to the environmental screening beyond the TMD layer always results in a dominant BGR term.

We stress that our findings stem from the effect of dynamical dielectric screening on the bound exciton: the self-energy of a bound exciton is not the self-energy of a free electron in the conduction band plus that of a free hole in the valence band. In a bound pair, the bandgap energy introduces a relative phase exp($iE_gt/\hbar$) between the electron and hole components, and therefore, at least one of these components is influenced by the high-frequency dielectric constant (Supplementary Theoretical Methods). This subtle but key detail is lost if one considers only the self-energy of a free particle, which is influenced by the static dielectric constant because the reference energy level in this case is the edge of the relevant energy band. This has important experimental consequences: if $\varepsilon (0)$ is very different from $\varepsilon (\infty)$, measurements of single-particle electronic bandgap such as by e.g. ARPES or Scanning Tunneling Spectroscopy \cite{ugeda_giant_2014} provide incorrect results to derive the BGR of a bound electron-hole pair.

Our results are also consistent with works that utilized graphene as a screening layer for the Coulomb engineering of excitons in TMDs \cite{tebbe_tailoring_2023, raja_coulomb_2017}. Unlike in our case, where strongly polar oxides result in extremely high $\varepsilon ({0})$, the large carrier mobility in graphene results in a very effective electronic screening, which in turn leads to a low $r_{2}$ and redshifting X$^{0}$ in TMDs.
Our results also indicate that the exciton self-energy and the exciton binding energy can be individually controlled by selecting screening materials with different $r_{2}$ values. Achieving the highest exciton energy difference at a dielectric heterojunction requires maximizing the $\Delta r_{2}$ between the different dielectric materials. 

\subsection*{\label{sec:trion}Effect of the dielectric screening on short-range Coulomb interactions}

\begin{figure}
\includegraphics[width=0.5\textwidth]{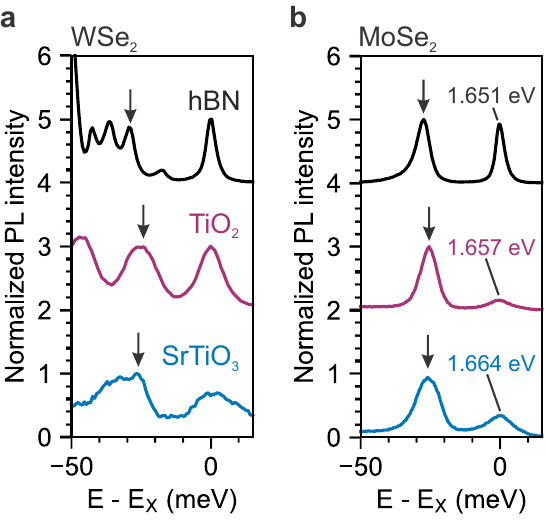}
\caption{\label{fig:trion}\textbf{Effect of the dielectric environment on the trion binding energy. }
\newline
\textbf{a}, PL spectra of monolayer WSe$_{2}$ on hBN, TiO$_{2}$, and SrTiO$_{3}$ in the electron doping regime. $\mathrm{E_{X}}$ is taken as the origin of the energy axis. In WSe$_{2}$, the negative trions are exchange-split. In each spectrum, X$^{-}_{intra}$ is indicated by a black arrow. \textbf{b}, PL spectra of monolayer MoSe$_{2}$ on hBN, TiO$_{2}$, and SrTiO$_{3}$ in the electron doping regime. $\mathrm{E_{X}}$ is taken as the origin of the energy axis, and it is indicated on the plots. X$^{-}$ is indicated by a black arrow in each spectrum.
}
\end{figure}

To understand dielectric screening effects on many-body complexes beyond excitons, we also experimentally investigate the behavior of the trion. Figure \ref{fig:trion}a shows the PL spectra of monolayer WSe$_{2}$ for the hBN, TiO$_{2}$, and SrTiO$_{3}$ samples in the electron doping regime, but close to charge neutrality (the X$^{0}$ and negative trions intensities are comparable) to minimize energy shifts from charge doping. The exciton resonance X$^{0}$ of each sample is taken as the origin of the energy axis to allow for a direct comparison of the trion binding energy across the different dielectric configurations. The negatively-charged intravalley trion X$^{-}_{\mathrm{intra}}$ shows only a weak dependence on $r_{2}$. Its binding energy starts at $\sim30$ meV in the hBN sample, drops to $\sim24$ meV in the TiO$_{2}$ sample, and rises to $\sim27$ meV in the SrTiO$_{3}$ sample. The non-monotonic behavior may be attributed to residual energy shifts from inconsistencies in charge doping among samples. To investigate the same effect in a material with spectrally well-separated resonances, we also study the X$^{-}$ binding energy in monolayer MoSe$_{2}$. Figure \ref{fig:trion}b shows the PL spectra of monolayer MoSe$_{2}$ for the hBN, TiO$_{2}$, and SrTiO$_{3}$ samples in the electron doping regime. We first note that X$^{0}$ in MoSe$_{2}$ also experiences a blueshift of up to $13$ meV at higher values of $r_{2}$. However, we do not observe any meaningful X$^{-}$ binding energy dependence on $r_{2}$, with the change being of the order of only a few meV.

The weak sensitivity of the trion binding energy in WSe$_{2}$ and MoSe$_{2}$ to $r_{2}$, together with the conservation of many of the excitonic features at extreme $r_{2}$ values, suggests that the formation of trions and other excitonic few-body complexes is only weakly affected by the static dielectric constant. At large distances, the interaction between a neutral exciton and an extra charge is dipolar in nature and, thus, has a relatively fast decay ($V(r)\sim 1/r^{2}$). Consequently, the binding energy of few-body complexes such as the trion is governed by short-range interparticle interactions, which are not sensitive to static screening \cite{van_tuan_coulomb_2018}.

\subsection*{\label{sec:outlook}Conclusion}

Coulomb interactions in atomically thin semiconductors coupled to polar oxides require a physical description beyond the static dielectric constant approximation, breaking the monolithic picture of exciton binding energy change and BGR as effects governed by the same type of screening, andrevealing a nuanced interplay of phenomena with distinct frequency dependence.
Our results offer new avenues to study and manipulate many-body interactions, and provide the necessary physical understanding to predict exciton behaviour when integrating TMDs and functional oxides. A natural consequence of our work would be to couple states with built-in electrical dipole with polar oxides, such as Janus TMDs\cite{petric_nonlinear_2023}, or tune of interlayer and moiré excitons via the dielectric environment. Using excitonic resonances as sensors for charge ordering could now provide deeper insights into correlated states. An exciting direction would be to explore the tuning of long-range interactions in strongly correlated systems, for example, in systems realizing the extended Hubbard model. This may allow the realization of currently inaccessible many-body phases, including interaction-induced Chern insulators and quantum spin liquids \cite{wu_hubbard_2018,pan_quantum_2020}. Finally, enabling the deterministic fabrication of dielectric superlattices could unlock the study of strongly correlated physics in artificial solid-state crystals and quasicrystals \cite{forsythe_band_2018}.

\section*{\label{sec:Methods}Methods}

\subsection*{\label{sec:methods:sample}Sample preparation}
All TMD, Gr, and hBN flakes were mechanically exfoliated from bulk crystals on SiO$_{2}$ substrates. The flakes were selected based on their optical contrast, shape, and cleanliness. The devices were assembled via the dry-transfer technique using polycarbonate films \cite{purdie_cleaning_2018} for the hBN and TiO$_{2}$ devices and using polypropylene carbonate \cite{wang_one-dimensional_2013} for the SrTiO$_{3}$ devices. Contacts to the respective layers were patterned using optical lithography and electron beam evaporation (Cr/Au $5/100$ nm). 
Single-crystal substrates of $(001)$ TiO$_{2}$ and $(100)$ SrTiO$_{3}$ were acquired from Shinkosha Co., Ltd.

\subsection*{\label{sec:methods:spectroscopy}Optical spectroscopy}
The optical measurements were performed in a variable-temperature helium flow cryostat with a confocal microscope in reflection geometry. For the PL measurements, $633$ nm / $532$ nm continuous wave laser sources were used for the excitation. The laser beam was focused onto the sample using an objective with a numerical aperture of $0.75$, yielding an excitation spot size of around $1$ $\mu$m. A pinhole was used as a spatial filter to obtain a diffraction-limited collection spot. The collected light is dispersed using a grating monochromator and detected on a CCD sensor array. The laser light was filtered using a $650/550$ nm short-pass filter. For reflection contrast spectroscopy, thermal light from a tungsten halogen light source was used for excitation. The gate voltage in the gate-tunable measurements was controlled using a Keithley $2400$ source meter.
Unless otherwise specified, all measurements presented here were performed at $10$ K.
A close-cycle optical cryostat in reflection geometry (Attocube, attoDRY800) with variable-temperature capability was used to perform the temperature-dependent measurements presented in the Supplementary.

\section*{\label{sec:data_availability}Data availability}
The datasets generated and analyzed during the current study are available from the corresponding authors upon reasonable request.

\bibliography{dielectric}

\begin{thebibliography}{46}%
\makeatletter
\providecommand \@ifxundefined [1]{%
 \@ifx{#1\undefined}
}%
\providecommand \@ifnum [1]{%
 \ifnum #1\expandafter \@firstoftwo
 \else \expandafter \@secondoftwo
 \fi
}%
\providecommand \@ifx [1]{%
 \ifx #1\expandafter \@firstoftwo
 \else \expandafter \@secondoftwo
 \fi
}%
\providecommand \natexlab [1]{#1}%
\providecommand \enquote  [1]{``#1''}%
\providecommand \bibnamefont  [1]{#1}%
\providecommand \bibfnamefont [1]{#1}%
\providecommand \citenamefont [1]{#1}%
\providecommand \href@noop [0]{\@secondoftwo}%
\providecommand \href [0]{\begingroup \@sanitize@url \@href}%
\providecommand \@href[1]{\@@startlink{#1}\@@href}%
\providecommand \@@href[1]{\endgroup#1\@@endlink}%
\providecommand \@sanitize@url [0]{\catcode `\\12\catcode `\$12\catcode `\&12\catcode `\#12\catcode `\^12\catcode `\_12\catcode `\%12\relax}%
\providecommand \@@startlink[1]{}%
\providecommand \@@endlink[0]{}%
\providecommand \url  [0]{\begingroup\@sanitize@url \@url }%
\providecommand \@url [1]{\endgroup\@href {#1}{\urlprefix }}%
\providecommand \urlprefix  [0]{URL }%
\providecommand \Eprint [0]{\href }%
\providecommand \doibase [0]{https://doi.org/}%
\providecommand \selectlanguage [0]{\@gobble}%
\providecommand \bibinfo  [0]{\@secondoftwo}%
\providecommand \bibfield  [0]{\@secondoftwo}%
\providecommand \translation [1]{[#1]}%
\providecommand \BibitemOpen [0]{}%
\providecommand \bibitemStop [0]{}%
\providecommand \bibitemNoStop [0]{.\EOS\space}%
\providecommand \EOS [0]{\spacefactor3000\relax}%
\providecommand \BibitemShut  [1]{\csname bibitem#1\endcsname}%
\let\auto@bib@innerbib\@empty
\bibitem [{\citenamefont {Snoke}(2019)}]{snoke_solid_2019}%
  \BibitemOpen
  \bibfield  {author} {\bibinfo {author} {\bibfnamefont {D.~W.}\ \bibnamefont {Snoke}},\ }\href@noop {} {\emph {\bibinfo {title} {{Solid State Physics: Essential Concepts}}}},\ \bibinfo {edition} {2nd}\ ed.\ (\bibinfo  {publisher} {Cambridge University Press},\ \bibinfo {address} {Cambridge, United Kingdom ; New York, NY},\ \bibinfo {year} {2019})\BibitemShut {NoStop}%
\bibitem [{\citenamefont {Yankowitz}\ \emph {et~al.}(2019)\citenamefont {Yankowitz}, \citenamefont {Ma}, \citenamefont {{Jarillo-Herrero}},\ and\ \citenamefont {LeRoy}}]{yankowitz_van_2019}%
  \BibitemOpen
  \bibfield  {author} {\bibinfo {author} {\bibfnamefont {M.}~\bibnamefont {Yankowitz}}, \bibinfo {author} {\bibfnamefont {Q.}~\bibnamefont {Ma}}, \bibinfo {author} {\bibfnamefont {P.}~\bibnamefont {{Jarillo-Herrero}}},\ and\ \bibinfo {author} {\bibfnamefont {B.~J.}\ \bibnamefont {LeRoy}},\ }\bibfield  {title} {\bibinfo {title} {Van der {{Waals}} heterostructures combining graphene and hexagonal boron nitride},\ }\href {https://doi.org/10.1038/s42254-018-0016-0} {\bibfield  {journal} {\bibinfo  {journal} {Nature Reviews Physics}\ }\textbf {\bibinfo {volume} {1}},\ \bibinfo {pages} {112} (\bibinfo {year} {2019})}\BibitemShut {NoStop}%
\bibitem [{\citenamefont {Mak}\ and\ \citenamefont {Shan}(2022)}]{mak_semiconductor_2022}%
  \BibitemOpen
  \bibfield  {author} {\bibinfo {author} {\bibfnamefont {K.~F.}\ \bibnamefont {Mak}}\ and\ \bibinfo {author} {\bibfnamefont {J.}~\bibnamefont {Shan}},\ }\bibfield  {title} {\bibinfo {title} {Semiconductor moir{\'e} materials},\ }\href {https://doi.org/10.1038/s41565-022-01165-6} {\bibfield  {journal} {\bibinfo  {journal} {Nature Nanotechnology}\ }\textbf {\bibinfo {volume} {17}},\ \bibinfo {pages} {686} (\bibinfo {year} {2022})}\BibitemShut {NoStop}%
\bibitem [{\citenamefont {Wilson}\ \emph {et~al.}(2021)\citenamefont {Wilson}, \citenamefont {Yao}, \citenamefont {Shan},\ and\ \citenamefont {Xu}}]{wilson_excitons_2021}%
  \BibitemOpen
  \bibfield  {author} {\bibinfo {author} {\bibfnamefont {N.~P.}\ \bibnamefont {Wilson}}, \bibinfo {author} {\bibfnamefont {W.}~\bibnamefont {Yao}}, \bibinfo {author} {\bibfnamefont {J.}~\bibnamefont {Shan}},\ and\ \bibinfo {author} {\bibfnamefont {X.}~\bibnamefont {Xu}},\ }\bibfield  {title} {\bibinfo {title} {Excitons and emergent quantum phenomena in stacked {{2D}} semiconductors},\ }\href {https://doi.org/10.1038/s41586-021-03979-1} {\bibfield  {journal} {\bibinfo  {journal} {Nature}\ }\textbf {\bibinfo {volume} {599}},\ \bibinfo {pages} {383} (\bibinfo {year} {2021})}\BibitemShut {NoStop}%
\bibitem [{\citenamefont {Regan}\ \emph {et~al.}(2022)\citenamefont {Regan}, \citenamefont {Wang}, \citenamefont {Paik}, \citenamefont {Zeng}, \citenamefont {Zhang}, \citenamefont {Zhu}, \citenamefont {MacDonald}, \citenamefont {Deng},\ and\ \citenamefont {Wang}}]{regan_emerging_2022}%
  \BibitemOpen
  \bibfield  {author} {\bibinfo {author} {\bibfnamefont {E.~C.}\ \bibnamefont {Regan}}, \bibinfo {author} {\bibfnamefont {D.}~\bibnamefont {Wang}}, \bibinfo {author} {\bibfnamefont {E.~Y.}\ \bibnamefont {Paik}}, \bibinfo {author} {\bibfnamefont {Y.}~\bibnamefont {Zeng}}, \bibinfo {author} {\bibfnamefont {L.}~\bibnamefont {Zhang}}, \bibinfo {author} {\bibfnamefont {J.}~\bibnamefont {Zhu}}, \bibinfo {author} {\bibfnamefont {A.~H.}\ \bibnamefont {MacDonald}}, \bibinfo {author} {\bibfnamefont {H.}~\bibnamefont {Deng}},\ and\ \bibinfo {author} {\bibfnamefont {F.}~\bibnamefont {Wang}},\ }\bibfield  {title} {\bibinfo {title} {Emerging exciton physics in transition metal dichalcogenide heterobilayers},\ }\href {https://doi.org/10.1038/s41578-022-00440-1} {\bibfield  {journal} {\bibinfo  {journal} {Nature Reviews Materials}\ }\textbf {\bibinfo {volume} {7}},\ \bibinfo {pages} {778} (\bibinfo {year} {2022})}\BibitemShut {NoStop}%
\bibitem [{\citenamefont {Montblanch}\ \emph {et~al.}(2023)\citenamefont {Montblanch}, \citenamefont {Barbone}, \citenamefont {Aharonovich}, \citenamefont {Atat{\"u}re},\ and\ \citenamefont {Ferrari}}]{montblanch_layered_2023}%
  \BibitemOpen
  \bibfield  {author} {\bibinfo {author} {\bibfnamefont {A.~R.-P.}\ \bibnamefont {Montblanch}}, \bibinfo {author} {\bibfnamefont {M.}~\bibnamefont {Barbone}}, \bibinfo {author} {\bibfnamefont {I.}~\bibnamefont {Aharonovich}}, \bibinfo {author} {\bibfnamefont {M.}~\bibnamefont {Atat{\"u}re}},\ and\ \bibinfo {author} {\bibfnamefont {A.~C.}\ \bibnamefont {Ferrari}},\ }\bibfield  {title} {\bibinfo {title} {Layered materials as a platform for quantum technologies},\ }\href {https://doi.org/10.1038/s41565-023-01354-x} {\bibfield  {journal} {\bibinfo  {journal} {Nature Nanotechnology}\ }\textbf {\bibinfo {volume} {18}},\ \bibinfo {pages} {555} (\bibinfo {year} {2023})}\BibitemShut {NoStop}%
\bibitem [{\citenamefont {Barbone}\ \emph {et~al.}(2018)\citenamefont {Barbone}, \citenamefont {Montblanch}, \citenamefont {Kara}, \citenamefont {{Palacios-Berraquero}}, \citenamefont {Cadore}, \citenamefont {De~Fazio}, \citenamefont {Pingault}, \citenamefont {Mostaani}, \citenamefont {Li}, \citenamefont {Chen}, \citenamefont {Watanabe}, \citenamefont {Taniguchi}, \citenamefont {Tongay}, \citenamefont {Wang}, \citenamefont {Ferrari},\ and\ \citenamefont {Atat{\"u}re}}]{barbone_charge-tuneable_2018}%
  \BibitemOpen
  \bibfield  {author} {\bibinfo {author} {\bibfnamefont {M.}~\bibnamefont {Barbone}}, \bibinfo {author} {\bibfnamefont {A.~R.-P.}\ \bibnamefont {Montblanch}}, \bibinfo {author} {\bibfnamefont {D.~M.}\ \bibnamefont {Kara}}, \bibinfo {author} {\bibfnamefont {C.}~\bibnamefont {{Palacios-Berraquero}}}, \bibinfo {author} {\bibfnamefont {A.~R.}\ \bibnamefont {Cadore}}, \bibinfo {author} {\bibfnamefont {D.}~\bibnamefont {De~Fazio}}, \bibinfo {author} {\bibfnamefont {B.}~\bibnamefont {Pingault}}, \bibinfo {author} {\bibfnamefont {E.}~\bibnamefont {Mostaani}}, \bibinfo {author} {\bibfnamefont {H.}~\bibnamefont {Li}}, \bibinfo {author} {\bibfnamefont {B.}~\bibnamefont {Chen}}, \bibinfo {author} {\bibfnamefont {K.}~\bibnamefont {Watanabe}}, \bibinfo {author} {\bibfnamefont {T.}~\bibnamefont {Taniguchi}}, \bibinfo {author} {\bibfnamefont {S.}~\bibnamefont {Tongay}}, \bibinfo {author} {\bibfnamefont {G.}~\bibnamefont {Wang}}, \bibinfo {author} {\bibfnamefont {A.~C.}\ \bibnamefont {Ferrari}},\ and\ \bibinfo {author}
  {\bibfnamefont {M.}~\bibnamefont {Atat{\"u}re}},\ }\bibfield  {title} {\bibinfo {title} {Charge-tuneable biexciton complexes in monolayer {{WSe$_{2}$}}},\ }\href {https://doi.org/10.1038/s41467-018-05632-4} {\bibfield  {journal} {\bibinfo  {journal} {Nature Communications}\ }\textbf {\bibinfo {volume} {9}},\ \bibinfo {pages} {3721} (\bibinfo {year} {2018})}\BibitemShut {NoStop}%
\bibitem [{\citenamefont {Sidler}\ \emph {et~al.}(2017)\citenamefont {Sidler}, \citenamefont {Back}, \citenamefont {Cotlet}, \citenamefont {Srivastava}, \citenamefont {Fink}, \citenamefont {Kroner}, \citenamefont {Demler},\ and\ \citenamefont {Imamoglu}}]{sidler_fermi_2017}%
  \BibitemOpen
  \bibfield  {author} {\bibinfo {author} {\bibfnamefont {M.}~\bibnamefont {Sidler}}, \bibinfo {author} {\bibfnamefont {P.}~\bibnamefont {Back}}, \bibinfo {author} {\bibfnamefont {O.}~\bibnamefont {Cotlet}}, \bibinfo {author} {\bibfnamefont {A.}~\bibnamefont {Srivastava}}, \bibinfo {author} {\bibfnamefont {T.}~\bibnamefont {Fink}}, \bibinfo {author} {\bibfnamefont {M.}~\bibnamefont {Kroner}}, \bibinfo {author} {\bibfnamefont {E.}~\bibnamefont {Demler}},\ and\ \bibinfo {author} {\bibfnamefont {A.}~\bibnamefont {Imamoglu}},\ }\bibfield  {title} {\bibinfo {title} {Fermi polaron-polaritons in charge-tunable atomically thin semiconductors},\ }\href {https://doi.org/10.1038/nphys3949} {\bibfield  {journal} {\bibinfo  {journal} {Nature Physics}\ }\textbf {\bibinfo {volume} {13}},\ \bibinfo {pages} {255} (\bibinfo {year} {2017})}\BibitemShut {NoStop}%
\bibitem [{\citenamefont {Van~Tuan}\ \emph {et~al.}(2022)\citenamefont {Van~Tuan}, \citenamefont {Shi}, \citenamefont {Xu}, \citenamefont {Crooker},\ and\ \citenamefont {Dery}}]{van_tuan_six-body_2022}%
  \BibitemOpen
  \bibfield  {author} {\bibinfo {author} {\bibfnamefont {D.}~\bibnamefont {Van~Tuan}}, \bibinfo {author} {\bibfnamefont {S.-F.}\ \bibnamefont {Shi}}, \bibinfo {author} {\bibfnamefont {X.}~\bibnamefont {Xu}}, \bibinfo {author} {\bibfnamefont {S.~A.}\ \bibnamefont {Crooker}},\ and\ \bibinfo {author} {\bibfnamefont {H.}~\bibnamefont {Dery}},\ }\bibfield  {title} {\bibinfo {title} {Six-body and eight-body exciton states in monolayer {{WSe$_{2}$}}},\ }\href {https://doi.org/10.1103/PhysRevLett.129.076801} {\bibfield  {journal} {\bibinfo  {journal} {Physical Review Letters}\ }\textbf {\bibinfo {volume} {129}},\ \bibinfo {pages} {076801} (\bibinfo {year} {2022})}\BibitemShut {NoStop}%
\bibitem [{\citenamefont {Smole{\'n}ski}\ \emph {et~al.}(2021)\citenamefont {Smole{\'n}ski}, \citenamefont {Dolgirev}, \citenamefont {Kuhlenkamp}, \citenamefont {Popert}, \citenamefont {Shimazaki}, \citenamefont {Back}, \citenamefont {Lu}, \citenamefont {Kroner}, \citenamefont {Watanabe}, \citenamefont {Taniguchi}, \citenamefont {Esterlis}, \citenamefont {Demler},\ and\ \citenamefont {Imamo{\u g}lu}}]{smolenski_signatures_2021}%
  \BibitemOpen
  \bibfield  {author} {\bibinfo {author} {\bibfnamefont {T.}~\bibnamefont {Smole{\'n}ski}}, \bibinfo {author} {\bibfnamefont {P.~E.}\ \bibnamefont {Dolgirev}}, \bibinfo {author} {\bibfnamefont {C.}~\bibnamefont {Kuhlenkamp}}, \bibinfo {author} {\bibfnamefont {A.}~\bibnamefont {Popert}}, \bibinfo {author} {\bibfnamefont {Y.}~\bibnamefont {Shimazaki}}, \bibinfo {author} {\bibfnamefont {P.}~\bibnamefont {Back}}, \bibinfo {author} {\bibfnamefont {X.}~\bibnamefont {Lu}}, \bibinfo {author} {\bibfnamefont {M.}~\bibnamefont {Kroner}}, \bibinfo {author} {\bibfnamefont {K.}~\bibnamefont {Watanabe}}, \bibinfo {author} {\bibfnamefont {T.}~\bibnamefont {Taniguchi}}, \bibinfo {author} {\bibfnamefont {I.}~\bibnamefont {Esterlis}}, \bibinfo {author} {\bibfnamefont {E.}~\bibnamefont {Demler}},\ and\ \bibinfo {author} {\bibfnamefont {A.}~\bibnamefont {Imamo{\u g}lu}},\ }\bibfield  {title} {\bibinfo {title} {Signatures of {{Wigner}} crystal of electrons in a monolayer semiconductor},\ }\href
  {https://doi.org/10.1038/s41586-021-03590-4} {\bibfield  {journal} {\bibinfo  {journal} {Nature}\ }\textbf {\bibinfo {volume} {595}},\ \bibinfo {pages} {53} (\bibinfo {year} {2021})}\BibitemShut {NoStop}%
\bibitem [{\citenamefont {Tang}\ \emph {et~al.}(2020)\citenamefont {Tang}, \citenamefont {Li}, \citenamefont {Li}, \citenamefont {Xu}, \citenamefont {Liu}, \citenamefont {Barmak}, \citenamefont {Watanabe}, \citenamefont {Taniguchi}, \citenamefont {MacDonald}, \citenamefont {Shan},\ and\ \citenamefont {Mak}}]{tang_simulation_2020}%
  \BibitemOpen
  \bibfield  {author} {\bibinfo {author} {\bibfnamefont {Y.}~\bibnamefont {Tang}}, \bibinfo {author} {\bibfnamefont {L.}~\bibnamefont {Li}}, \bibinfo {author} {\bibfnamefont {T.}~\bibnamefont {Li}}, \bibinfo {author} {\bibfnamefont {Y.}~\bibnamefont {Xu}}, \bibinfo {author} {\bibfnamefont {S.}~\bibnamefont {Liu}}, \bibinfo {author} {\bibfnamefont {K.}~\bibnamefont {Barmak}}, \bibinfo {author} {\bibfnamefont {K.}~\bibnamefont {Watanabe}}, \bibinfo {author} {\bibfnamefont {T.}~\bibnamefont {Taniguchi}}, \bibinfo {author} {\bibfnamefont {A.~H.}\ \bibnamefont {MacDonald}}, \bibinfo {author} {\bibfnamefont {J.}~\bibnamefont {Shan}},\ and\ \bibinfo {author} {\bibfnamefont {K.~F.}\ \bibnamefont {Mak}},\ }\bibfield  {title} {\bibinfo {title} {Simulation of {{Hubbard}} model physics in {{WSe$_{2}$}}/{{WS$_{2}$}} moir{\'e} superlattices},\ }\href {https://doi.org/10.1038/s41586-020-2085-3} {\bibfield  {journal} {\bibinfo  {journal} {Nature}\ }\textbf {\bibinfo {volume} {579}},\ \bibinfo {pages} {353} (\bibinfo {year}
  {2020})}\BibitemShut {NoStop}%
\bibitem [{\citenamefont {Xu}\ \emph {et~al.}(2020)\citenamefont {Xu}, \citenamefont {Liu}, \citenamefont {Rhodes}, \citenamefont {Watanabe}, \citenamefont {Taniguchi}, \citenamefont {Hone}, \citenamefont {Elser}, \citenamefont {Mak},\ and\ \citenamefont {Shan}}]{xu_correlated_2020}%
  \BibitemOpen
  \bibfield  {author} {\bibinfo {author} {\bibfnamefont {Y.}~\bibnamefont {Xu}}, \bibinfo {author} {\bibfnamefont {S.}~\bibnamefont {Liu}}, \bibinfo {author} {\bibfnamefont {D.~A.}\ \bibnamefont {Rhodes}}, \bibinfo {author} {\bibfnamefont {K.}~\bibnamefont {Watanabe}}, \bibinfo {author} {\bibfnamefont {T.}~\bibnamefont {Taniguchi}}, \bibinfo {author} {\bibfnamefont {J.}~\bibnamefont {Hone}}, \bibinfo {author} {\bibfnamefont {V.}~\bibnamefont {Elser}}, \bibinfo {author} {\bibfnamefont {K.~F.}\ \bibnamefont {Mak}},\ and\ \bibinfo {author} {\bibfnamefont {J.}~\bibnamefont {Shan}},\ }\bibfield  {title} {\bibinfo {title} {Correlated insulating states at fractional fillings of moir{\'e} superlattices},\ }\href {https://doi.org/10.1038/s41586-020-2868-6} {\bibfield  {journal} {\bibinfo  {journal} {Nature}\ }\textbf {\bibinfo {volume} {587}},\ \bibinfo {pages} {214} (\bibinfo {year} {2020})}\BibitemShut {NoStop}%
\bibitem [{\citenamefont {Cao}\ \emph {et~al.}(2018)\citenamefont {Cao}, \citenamefont {Fatemi}, \citenamefont {Fang}, \citenamefont {Watanabe}, \citenamefont {Taniguchi}, \citenamefont {Kaxiras},\ and\ \citenamefont {{Jarillo-Herrero}}}]{cao_unconventional_2018}%
  \BibitemOpen
  \bibfield  {author} {\bibinfo {author} {\bibfnamefont {Y.}~\bibnamefont {Cao}}, \bibinfo {author} {\bibfnamefont {V.}~\bibnamefont {Fatemi}}, \bibinfo {author} {\bibfnamefont {S.}~\bibnamefont {Fang}}, \bibinfo {author} {\bibfnamefont {K.}~\bibnamefont {Watanabe}}, \bibinfo {author} {\bibfnamefont {T.}~\bibnamefont {Taniguchi}}, \bibinfo {author} {\bibfnamefont {E.}~\bibnamefont {Kaxiras}},\ and\ \bibinfo {author} {\bibfnamefont {P.}~\bibnamefont {{Jarillo-Herrero}}},\ }\bibfield  {title} {\bibinfo {title} {Unconventional superconductivity in magic-angle graphene superlattices},\ }\href {https://doi.org/10.1038/nature26160} {\bibfield  {journal} {\bibinfo  {journal} {Nature}\ }\textbf {\bibinfo {volume} {556}},\ \bibinfo {pages} {43} (\bibinfo {year} {2018})}\BibitemShut {NoStop}%
\bibitem [{\citenamefont {Chen}\ \emph {et~al.}(2020)\citenamefont {Chen}, \citenamefont {Sharpe}, \citenamefont {Fox}, \citenamefont {Zhang}, \citenamefont {Wang}, \citenamefont {Jiang}, \citenamefont {Lyu}, \citenamefont {Li}, \citenamefont {Watanabe}, \citenamefont {Taniguchi}, \citenamefont {Shi}, \citenamefont {Senthil}, \citenamefont {{Goldhaber-Gordon}}, \citenamefont {Zhang},\ and\ \citenamefont {Wang}}]{chen_tunable_2020}%
  \BibitemOpen
  \bibfield  {author} {\bibinfo {author} {\bibfnamefont {G.}~\bibnamefont {Chen}}, \bibinfo {author} {\bibfnamefont {A.~L.}\ \bibnamefont {Sharpe}}, \bibinfo {author} {\bibfnamefont {E.~J.}\ \bibnamefont {Fox}}, \bibinfo {author} {\bibfnamefont {Y.-H.}\ \bibnamefont {Zhang}}, \bibinfo {author} {\bibfnamefont {S.}~\bibnamefont {Wang}}, \bibinfo {author} {\bibfnamefont {L.}~\bibnamefont {Jiang}}, \bibinfo {author} {\bibfnamefont {B.}~\bibnamefont {Lyu}}, \bibinfo {author} {\bibfnamefont {H.}~\bibnamefont {Li}}, \bibinfo {author} {\bibfnamefont {K.}~\bibnamefont {Watanabe}}, \bibinfo {author} {\bibfnamefont {T.}~\bibnamefont {Taniguchi}}, \bibinfo {author} {\bibfnamefont {Z.}~\bibnamefont {Shi}}, \bibinfo {author} {\bibfnamefont {T.}~\bibnamefont {Senthil}}, \bibinfo {author} {\bibfnamefont {D.}~\bibnamefont {{Goldhaber-Gordon}}}, \bibinfo {author} {\bibfnamefont {Y.}~\bibnamefont {Zhang}},\ and\ \bibinfo {author} {\bibfnamefont {F.}~\bibnamefont {Wang}},\ }\bibfield  {title} {\bibinfo {title} {Tunable
  correlated {{Chern}} insulator and ferromagnetism in a moir{\'e} superlattice},\ }\href {https://doi.org/10.1038/s41586-020-2049-7} {\bibfield  {journal} {\bibinfo  {journal} {Nature}\ }\textbf {\bibinfo {volume} {579}},\ \bibinfo {pages} {56} (\bibinfo {year} {2020})}\BibitemShut {NoStop}%
\bibitem [{\citenamefont {Rytova}(2020)}]{rytova_screened_2020}%
  \BibitemOpen
  \bibfield  {author} {\bibinfo {author} {\bibfnamefont {N.~S.}\ \bibnamefont {Rytova}},\ }\href {https://doi.org/10.48550/arXiv.1806.00976} {\bibinfo {title} {Screened potential of a point charge in a thin film}} (\bibinfo {year} {2020}),\ \Eprint {https://arxiv.org/abs/1806.00976} {arXiv:1806.00976} \BibitemShut {NoStop}%
\bibitem [{\citenamefont {Keldysh}(1979)}]{keldysh_coulomb_1979}%
  \BibitemOpen
  \bibfield  {author} {\bibinfo {author} {\bibfnamefont {L.~V.}\ \bibnamefont {Keldysh}},\ }\bibfield  {title} {\bibinfo {title} {Coulomb interaction in thin semiconductor and semimetal films},\ }\href {https://doi.org/10.1142/9789811279461_0024} {\bibfield  {journal} {\bibinfo  {journal} {JETP Letters}\ }\textbf {\bibinfo {volume} {29}},\ \bibinfo {pages} {658} (\bibinfo {year} {1979})}\BibitemShut {NoStop}%
\bibitem [{\citenamefont {Cudazzo}\ \emph {et~al.}(2011)\citenamefont {Cudazzo}, \citenamefont {Tokatly},\ and\ \citenamefont {Rubio}}]{cudazzo_dielectric_2011}%
  \BibitemOpen
  \bibfield  {author} {\bibinfo {author} {\bibfnamefont {P.}~\bibnamefont {Cudazzo}}, \bibinfo {author} {\bibfnamefont {I.~V.}\ \bibnamefont {Tokatly}},\ and\ \bibinfo {author} {\bibfnamefont {A.}~\bibnamefont {Rubio}},\ }\bibfield  {title} {\bibinfo {title} {Dielectric screening in two-dimensional insulators: {{Implications}} for excitonic and impurity states in graphane},\ }\href {https://doi.org/10.1103/PhysRevB.84.085406} {\bibfield  {journal} {\bibinfo  {journal} {Physical Review B}\ }\textbf {\bibinfo {volume} {84}},\ \bibinfo {pages} {085406} (\bibinfo {year} {2011})}\BibitemShut {NoStop}%
\bibitem [{\citenamefont {Chernikov}\ \emph {et~al.}(2014)\citenamefont {Chernikov}, \citenamefont {Berkelbach}, \citenamefont {Hill}, \citenamefont {Rigosi}, \citenamefont {Li}, \citenamefont {Aslan}, \citenamefont {Reichman}, \citenamefont {Hybertsen},\ and\ \citenamefont {Heinz}}]{chernikov_exciton_2014}%
  \BibitemOpen
  \bibfield  {author} {\bibinfo {author} {\bibfnamefont {A.}~\bibnamefont {Chernikov}}, \bibinfo {author} {\bibfnamefont {T.~C.}\ \bibnamefont {Berkelbach}}, \bibinfo {author} {\bibfnamefont {H.~M.}\ \bibnamefont {Hill}}, \bibinfo {author} {\bibfnamefont {A.}~\bibnamefont {Rigosi}}, \bibinfo {author} {\bibfnamefont {Y.}~\bibnamefont {Li}}, \bibinfo {author} {\bibfnamefont {B.}~\bibnamefont {Aslan}}, \bibinfo {author} {\bibfnamefont {D.~R.}\ \bibnamefont {Reichman}}, \bibinfo {author} {\bibfnamefont {M.~S.}\ \bibnamefont {Hybertsen}},\ and\ \bibinfo {author} {\bibfnamefont {T.~F.}\ \bibnamefont {Heinz}},\ }\bibfield  {title} {\bibinfo {title} {Exciton binding energy and nonhydrogenic {{Rydberg}} series in monolayer {{WS$_{2}$}}},\ }\href {https://doi.org/10.1103/PhysRevLett.113.076802} {\bibfield  {journal} {\bibinfo  {journal} {Physical Review Letters}\ }\textbf {\bibinfo {volume} {113}},\ \bibinfo {pages} {076802} (\bibinfo {year} {2014})}\BibitemShut {NoStop}%
\bibitem [{\citenamefont {Ugeda}\ \emph {et~al.}(2014)\citenamefont {Ugeda}, \citenamefont {Bradley}, \citenamefont {Shi}, \citenamefont {{da Jornada}}, \citenamefont {Zhang}, \citenamefont {Qiu}, \citenamefont {Ruan}, \citenamefont {Mo}, \citenamefont {Hussain}, \citenamefont {Shen}, \citenamefont {Wang}, \citenamefont {Louie},\ and\ \citenamefont {Crommie}}]{ugeda_giant_2014}%
  \BibitemOpen
  \bibfield  {author} {\bibinfo {author} {\bibfnamefont {M.~M.}\ \bibnamefont {Ugeda}}, \bibinfo {author} {\bibfnamefont {A.~J.}\ \bibnamefont {Bradley}}, \bibinfo {author} {\bibfnamefont {S.-F.}\ \bibnamefont {Shi}}, \bibinfo {author} {\bibfnamefont {F.~H.}\ \bibnamefont {{da Jornada}}}, \bibinfo {author} {\bibfnamefont {Y.}~\bibnamefont {Zhang}}, \bibinfo {author} {\bibfnamefont {D.~Y.}\ \bibnamefont {Qiu}}, \bibinfo {author} {\bibfnamefont {W.}~\bibnamefont {Ruan}}, \bibinfo {author} {\bibfnamefont {S.-K.}\ \bibnamefont {Mo}}, \bibinfo {author} {\bibfnamefont {Z.}~\bibnamefont {Hussain}}, \bibinfo {author} {\bibfnamefont {Z.-X.}\ \bibnamefont {Shen}}, \bibinfo {author} {\bibfnamefont {F.}~\bibnamefont {Wang}}, \bibinfo {author} {\bibfnamefont {S.~G.}\ \bibnamefont {Louie}},\ and\ \bibinfo {author} {\bibfnamefont {M.~F.}\ \bibnamefont {Crommie}},\ }\bibfield  {title} {\bibinfo {title} {Giant bandgap renormalization and excitonic effects in a monolayer transition metal dichalcogenide semiconductor},\ }\href
  {https://doi.org/10.1038/nmat4061} {\bibfield  {journal} {\bibinfo  {journal} {Nature Materials}\ }\textbf {\bibinfo {volume} {13}},\ \bibinfo {pages} {1091} (\bibinfo {year} {2014})}\BibitemShut {NoStop}%
\bibitem [{\citenamefont {Raja}\ \emph {et~al.}(2017)\citenamefont {Raja}, \citenamefont {Chaves}, \citenamefont {Yu}, \citenamefont {Arefe}, \citenamefont {Hill}, \citenamefont {Rigosi}, \citenamefont {Berkelbach}, \citenamefont {Nagler}, \citenamefont {Sch{\"u}ller}, \citenamefont {Korn}, \citenamefont {Nuckolls}, \citenamefont {Hone}, \citenamefont {Brus}, \citenamefont {Heinz}, \citenamefont {Reichman},\ and\ \citenamefont {Chernikov}}]{raja_coulomb_2017}%
  \BibitemOpen
  \bibfield  {author} {\bibinfo {author} {\bibfnamefont {A.}~\bibnamefont {Raja}}, \bibinfo {author} {\bibfnamefont {A.}~\bibnamefont {Chaves}}, \bibinfo {author} {\bibfnamefont {J.}~\bibnamefont {Yu}}, \bibinfo {author} {\bibfnamefont {G.}~\bibnamefont {Arefe}}, \bibinfo {author} {\bibfnamefont {H.~M.}\ \bibnamefont {Hill}}, \bibinfo {author} {\bibfnamefont {A.~F.}\ \bibnamefont {Rigosi}}, \bibinfo {author} {\bibfnamefont {T.~C.}\ \bibnamefont {Berkelbach}}, \bibinfo {author} {\bibfnamefont {P.}~\bibnamefont {Nagler}}, \bibinfo {author} {\bibfnamefont {C.}~\bibnamefont {Sch{\"u}ller}}, \bibinfo {author} {\bibfnamefont {T.}~\bibnamefont {Korn}}, \bibinfo {author} {\bibfnamefont {C.}~\bibnamefont {Nuckolls}}, \bibinfo {author} {\bibfnamefont {J.}~\bibnamefont {Hone}}, \bibinfo {author} {\bibfnamefont {L.~E.}\ \bibnamefont {Brus}}, \bibinfo {author} {\bibfnamefont {T.~F.}\ \bibnamefont {Heinz}}, \bibinfo {author} {\bibfnamefont {D.~R.}\ \bibnamefont {Reichman}},\ and\ \bibinfo {author} {\bibfnamefont
  {A.}~\bibnamefont {Chernikov}},\ }\bibfield  {title} {\bibinfo {title} {Coulomb engineering of the bandgap and excitons in two-dimensional materials},\ }\href {https://doi.org/10.1038/ncomms15251} {\bibfield  {journal} {\bibinfo  {journal} {Nature Communications}\ }\textbf {\bibinfo {volume} {8}},\ \bibinfo {pages} {15251} (\bibinfo {year} {2017})}\BibitemShut {NoStop}%
\bibitem [{\citenamefont {Forsythe}\ \emph {et~al.}(2018)\citenamefont {Forsythe}, \citenamefont {Zhou}, \citenamefont {Watanabe}, \citenamefont {Taniguchi}, \citenamefont {Pasupathy}, \citenamefont {Moon}, \citenamefont {Koshino}, \citenamefont {Kim},\ and\ \citenamefont {Dean}}]{forsythe_band_2018}%
  \BibitemOpen
  \bibfield  {author} {\bibinfo {author} {\bibfnamefont {C.}~\bibnamefont {Forsythe}}, \bibinfo {author} {\bibfnamefont {X.}~\bibnamefont {Zhou}}, \bibinfo {author} {\bibfnamefont {K.}~\bibnamefont {Watanabe}}, \bibinfo {author} {\bibfnamefont {T.}~\bibnamefont {Taniguchi}}, \bibinfo {author} {\bibfnamefont {A.}~\bibnamefont {Pasupathy}}, \bibinfo {author} {\bibfnamefont {P.}~\bibnamefont {Moon}}, \bibinfo {author} {\bibfnamefont {M.}~\bibnamefont {Koshino}}, \bibinfo {author} {\bibfnamefont {P.}~\bibnamefont {Kim}},\ and\ \bibinfo {author} {\bibfnamefont {C.~R.}\ \bibnamefont {Dean}},\ }\bibfield  {title} {\bibinfo {title} {Band structure engineering of {{2D}} materials using patterned dielectric superlattices},\ }\href {https://doi.org/10.1038/s41565-018-0138-7} {\bibfield  {journal} {\bibinfo  {journal} {Nature Nanotechnology}\ }\textbf {\bibinfo {volume} {13}},\ \bibinfo {pages} {566} (\bibinfo {year} {2018})}\BibitemShut {NoStop}%
\bibitem [{\citenamefont {Van~Tuan}\ \emph {et~al.}(2018)\citenamefont {Van~Tuan}, \citenamefont {Yang},\ and\ \citenamefont {Dery}}]{van_tuan_coulomb_2018}%
  \BibitemOpen
  \bibfield  {author} {\bibinfo {author} {\bibfnamefont {D.}~\bibnamefont {Van~Tuan}}, \bibinfo {author} {\bibfnamefont {M.}~\bibnamefont {Yang}},\ and\ \bibinfo {author} {\bibfnamefont {H.}~\bibnamefont {Dery}},\ }\bibfield  {title} {\bibinfo {title} {Coulomb interaction in monolayer transition-metal dichalcogenides},\ }\href {https://doi.org/10.1103/PhysRevB.98.125308} {\bibfield  {journal} {\bibinfo  {journal} {Physical Review B}\ }\textbf {\bibinfo {volume} {98}},\ \bibinfo {pages} {125308} (\bibinfo {year} {2018})}\BibitemShut {NoStop}%
\bibitem [{\citenamefont {Cho}\ and\ \citenamefont {Berkelbach}(2018)}]{cho_environmentally_2018}%
  \BibitemOpen
  \bibfield  {author} {\bibinfo {author} {\bibfnamefont {Y.}~\bibnamefont {Cho}}\ and\ \bibinfo {author} {\bibfnamefont {T.~C.}\ \bibnamefont {Berkelbach}},\ }\bibfield  {title} {\bibinfo {title} {Environmentally sensitive theory of electronic and optical transitions in atomically thin semiconductors},\ }\href {https://doi.org/10.1103/PhysRevB.97.041409} {\bibfield  {journal} {\bibinfo  {journal} {Physical Review B}\ }\textbf {\bibinfo {volume} {97}},\ \bibinfo {pages} {041409} (\bibinfo {year} {2018})}\BibitemShut {NoStop}%
\bibitem [{\citenamefont {Raja}\ \emph {et~al.}(2019)\citenamefont {Raja}, \citenamefont {Waldecker}, \citenamefont {Zipfel}, \citenamefont {Cho}, \citenamefont {Brem}, \citenamefont {Ziegler}, \citenamefont {Kulig}, \citenamefont {Taniguchi}, \citenamefont {Watanabe}, \citenamefont {Malic}, \citenamefont {Heinz}, \citenamefont {Berkelbach},\ and\ \citenamefont {Chernikov}}]{raja_dielectric_2019}%
  \BibitemOpen
  \bibfield  {author} {\bibinfo {author} {\bibfnamefont {A.}~\bibnamefont {Raja}}, \bibinfo {author} {\bibfnamefont {L.}~\bibnamefont {Waldecker}}, \bibinfo {author} {\bibfnamefont {J.}~\bibnamefont {Zipfel}}, \bibinfo {author} {\bibfnamefont {Y.}~\bibnamefont {Cho}}, \bibinfo {author} {\bibfnamefont {S.}~\bibnamefont {Brem}}, \bibinfo {author} {\bibfnamefont {J.~D.}\ \bibnamefont {Ziegler}}, \bibinfo {author} {\bibfnamefont {M.}~\bibnamefont {Kulig}}, \bibinfo {author} {\bibfnamefont {T.}~\bibnamefont {Taniguchi}}, \bibinfo {author} {\bibfnamefont {K.}~\bibnamefont {Watanabe}}, \bibinfo {author} {\bibfnamefont {E.}~\bibnamefont {Malic}}, \bibinfo {author} {\bibfnamefont {T.~F.}\ \bibnamefont {Heinz}}, \bibinfo {author} {\bibfnamefont {T.~C.}\ \bibnamefont {Berkelbach}},\ and\ \bibinfo {author} {\bibfnamefont {A.}~\bibnamefont {Chernikov}},\ }\bibfield  {title} {\bibinfo {title} {Dielectric disorder in two-dimensional materials},\ }\href {https://doi.org/10.1038/s41565-019-0520-0} {\bibfield  {journal}
  {\bibinfo  {journal} {Nature Nanotechnology}\ }\textbf {\bibinfo {volume} {14}},\ \bibinfo {pages} {832} (\bibinfo {year} {2019})}\BibitemShut {NoStop}%
\bibitem [{\citenamefont {Stier}\ \emph {et~al.}(2016)\citenamefont {Stier}, \citenamefont {Wilson}, \citenamefont {Clark}, \citenamefont {Xu},\ and\ \citenamefont {Crooker}}]{stier_probing_2016}%
  \BibitemOpen
  \bibfield  {author} {\bibinfo {author} {\bibfnamefont {A.~V.}\ \bibnamefont {Stier}}, \bibinfo {author} {\bibfnamefont {N.~P.}\ \bibnamefont {Wilson}}, \bibinfo {author} {\bibfnamefont {G.}~\bibnamefont {Clark}}, \bibinfo {author} {\bibfnamefont {X.}~\bibnamefont {Xu}},\ and\ \bibinfo {author} {\bibfnamefont {S.~A.}\ \bibnamefont {Crooker}},\ }\bibfield  {title} {\bibinfo {title} {Probing the influence of dielectric environment on excitons in monolayer {{WSe$_{2}$}}: Insight from high magnetic fields},\ }\href {https://doi.org/10.1021/acs.nanolett.6b03276} {\bibfield  {journal} {\bibinfo  {journal} {Nano Letters}\ }\textbf {\bibinfo {volume} {16}},\ \bibinfo {pages} {7054} (\bibinfo {year} {2016})}\BibitemShut {NoStop}%
\bibitem [{\citenamefont {Tebbe}\ \emph {et~al.}(2023)\citenamefont {Tebbe}, \citenamefont {Sch{\"u}tte}, \citenamefont {Watanabe}, \citenamefont {Taniguchi}, \citenamefont {Stampfer}, \citenamefont {Beschoten},\ and\ \citenamefont {Waldecker}}]{tebbe_tailoring_2023}%
  \BibitemOpen
  \bibfield  {author} {\bibinfo {author} {\bibfnamefont {D.}~\bibnamefont {Tebbe}}, \bibinfo {author} {\bibfnamefont {M.}~\bibnamefont {Sch{\"u}tte}}, \bibinfo {author} {\bibfnamefont {K.}~\bibnamefont {Watanabe}}, \bibinfo {author} {\bibfnamefont {T.}~\bibnamefont {Taniguchi}}, \bibinfo {author} {\bibfnamefont {C.}~\bibnamefont {Stampfer}}, \bibinfo {author} {\bibfnamefont {B.}~\bibnamefont {Beschoten}},\ and\ \bibinfo {author} {\bibfnamefont {L.}~\bibnamefont {Waldecker}},\ }\bibfield  {title} {\bibinfo {title} {Tailoring the dielectric screening in {{WS$_{2}$}}--graphene heterostructures},\ }\href {https://doi.org/10.1038/s41699-023-00394-0} {\bibfield  {journal} {\bibinfo  {journal} {npj 2D Materials and Applications}\ }\textbf {\bibinfo {volume} {7}},\ \bibinfo {pages} {1} (\bibinfo {year} {2023})}\BibitemShut {NoStop}%
\bibitem [{\citenamefont {Goryca}\ \emph {et~al.}(2019)\citenamefont {Goryca}, \citenamefont {Li}, \citenamefont {Stier}, \citenamefont {Taniguchi}, \citenamefont {Watanabe}, \citenamefont {Courtade}, \citenamefont {Shree}, \citenamefont {Robert}, \citenamefont {Urbaszek}, \citenamefont {Marie},\ and\ \citenamefont {Crooker}}]{goryca_revealing_2019}%
  \BibitemOpen
  \bibfield  {author} {\bibinfo {author} {\bibfnamefont {M.}~\bibnamefont {Goryca}}, \bibinfo {author} {\bibfnamefont {J.}~\bibnamefont {Li}}, \bibinfo {author} {\bibfnamefont {A.~V.}\ \bibnamefont {Stier}}, \bibinfo {author} {\bibfnamefont {T.}~\bibnamefont {Taniguchi}}, \bibinfo {author} {\bibfnamefont {K.}~\bibnamefont {Watanabe}}, \bibinfo {author} {\bibfnamefont {E.}~\bibnamefont {Courtade}}, \bibinfo {author} {\bibfnamefont {S.}~\bibnamefont {Shree}}, \bibinfo {author} {\bibfnamefont {C.}~\bibnamefont {Robert}}, \bibinfo {author} {\bibfnamefont {B.}~\bibnamefont {Urbaszek}}, \bibinfo {author} {\bibfnamefont {X.}~\bibnamefont {Marie}},\ and\ \bibinfo {author} {\bibfnamefont {S.~A.}\ \bibnamefont {Crooker}},\ }\bibfield  {title} {\bibinfo {title} {Revealing exciton masses and dielectric properties of monolayer semiconductors with high magnetic fields},\ }\href {https://doi.org/10.1038/s41467-019-12180-y} {\bibfield  {journal} {\bibinfo  {journal} {Nature Communications}\ }\textbf {\bibinfo {volume}
  {10}},\ \bibinfo {pages} {4172} (\bibinfo {year} {2019})}\BibitemShut {NoStop}%
\bibitem [{\citenamefont {Scharf}\ \emph {et~al.}(2019)\citenamefont {Scharf}, \citenamefont {Tuan}, \citenamefont {{\v Z}uti{\'c}},\ and\ \citenamefont {Dery}}]{scharf_dynamical_2019}%
  \BibitemOpen
  \bibfield  {author} {\bibinfo {author} {\bibfnamefont {B.}~\bibnamefont {Scharf}}, \bibinfo {author} {\bibfnamefont {D.~V.}\ \bibnamefont {Tuan}}, \bibinfo {author} {\bibfnamefont {I.}~\bibnamefont {{\v Z}uti{\'c}}},\ and\ \bibinfo {author} {\bibfnamefont {H.}~\bibnamefont {Dery}},\ }\bibfield  {title} {\bibinfo {title} {Dynamical screening in monolayer transition-metal dichalcogenides and its manifestations in the exciton spectrum},\ }\href {https://doi.org/10.1088/1361-648X/ab071f} {\bibfield  {journal} {\bibinfo  {journal} {Journal of Physics: Condensed Matter}\ }\textbf {\bibinfo {volume} {31}},\ \bibinfo {pages} {203001} (\bibinfo {year} {2019})}\BibitemShut {NoStop}%
\bibitem [{\citenamefont {Stier}\ \emph {et~al.}(2018)\citenamefont {Stier}, \citenamefont {Wilson}, \citenamefont {Velizhanin}, \citenamefont {Kono}, \citenamefont {Xu},\ and\ \citenamefont {Crooker}}]{stier_magnetooptics_2018}%
  \BibitemOpen
  \bibfield  {author} {\bibinfo {author} {\bibfnamefont {A.~V.}\ \bibnamefont {Stier}}, \bibinfo {author} {\bibfnamefont {N.~P.}\ \bibnamefont {Wilson}}, \bibinfo {author} {\bibfnamefont {K.~A.}\ \bibnamefont {Velizhanin}}, \bibinfo {author} {\bibfnamefont {J.}~\bibnamefont {Kono}}, \bibinfo {author} {\bibfnamefont {X.}~\bibnamefont {Xu}},\ and\ \bibinfo {author} {\bibfnamefont {S.~A.}\ \bibnamefont {Crooker}},\ }\bibfield  {title} {\bibinfo {title} {Magnetooptics of exciton {{Rydberg}} states in a monolayer semiconductor},\ }\href {https://doi.org/10.1103/PhysRevLett.120.057405} {\bibfield  {journal} {\bibinfo  {journal} {Physical Review Letters}\ }\textbf {\bibinfo {volume} {120}},\ \bibinfo {pages} {057405} (\bibinfo {year} {2018})}\BibitemShut {NoStop}%
\bibitem [{\citenamefont {He}\ \emph {et~al.}(2020)\citenamefont {He}, \citenamefont {Rivera}, \citenamefont {Van~Tuan}, \citenamefont {Wilson}, \citenamefont {Yang}, \citenamefont {Taniguchi}, \citenamefont {Watanabe}, \citenamefont {Yan}, \citenamefont {Mandrus}, \citenamefont {Yu}, \citenamefont {Dery}, \citenamefont {Yao},\ and\ \citenamefont {Xu}}]{he_valley_2020}%
  \BibitemOpen
  \bibfield  {author} {\bibinfo {author} {\bibfnamefont {M.}~\bibnamefont {He}}, \bibinfo {author} {\bibfnamefont {P.}~\bibnamefont {Rivera}}, \bibinfo {author} {\bibfnamefont {D.}~\bibnamefont {Van~Tuan}}, \bibinfo {author} {\bibfnamefont {N.~P.}\ \bibnamefont {Wilson}}, \bibinfo {author} {\bibfnamefont {M.}~\bibnamefont {Yang}}, \bibinfo {author} {\bibfnamefont {T.}~\bibnamefont {Taniguchi}}, \bibinfo {author} {\bibfnamefont {K.}~\bibnamefont {Watanabe}}, \bibinfo {author} {\bibfnamefont {J.}~\bibnamefont {Yan}}, \bibinfo {author} {\bibfnamefont {D.~G.}\ \bibnamefont {Mandrus}}, \bibinfo {author} {\bibfnamefont {H.}~\bibnamefont {Yu}}, \bibinfo {author} {\bibfnamefont {H.}~\bibnamefont {Dery}}, \bibinfo {author} {\bibfnamefont {W.}~\bibnamefont {Yao}},\ and\ \bibinfo {author} {\bibfnamefont {X.}~\bibnamefont {Xu}},\ }\bibfield  {title} {\bibinfo {title} {Valley phonons and exciton complexes in a monolayer semiconductor},\ }\href {https://doi.org/10.1038/s41467-020-14472-0} {\bibfield  {journal} {\bibinfo
  {journal} {Nature Communications}\ }\textbf {\bibinfo {volume} {11}},\ \bibinfo {pages} {618} (\bibinfo {year} {2020})}\BibitemShut {NoStop}%
\bibitem [{\citenamefont {Sch{\"o}che}\ \emph {et~al.}(2013)\citenamefont {Sch{\"o}che}, \citenamefont {Hofmann}, \citenamefont {Korlacki}, \citenamefont {Tiwald},\ and\ \citenamefont {Schubert}}]{schoche_infrared_2013}%
  \BibitemOpen
  \bibfield  {author} {\bibinfo {author} {\bibfnamefont {S.}~\bibnamefont {Sch{\"o}che}}, \bibinfo {author} {\bibfnamefont {T.}~\bibnamefont {Hofmann}}, \bibinfo {author} {\bibfnamefont {R.}~\bibnamefont {Korlacki}}, \bibinfo {author} {\bibfnamefont {T.~E.}\ \bibnamefont {Tiwald}},\ and\ \bibinfo {author} {\bibfnamefont {M.}~\bibnamefont {Schubert}},\ }\bibfield  {title} {\bibinfo {title} {Infrared dielectric anisotropy and phonon modes of rutile {{TiO$_{2}$}}},\ }\href {https://doi.org/10.1063/1.4802715} {\bibfield  {journal} {\bibinfo  {journal} {Journal of Applied Physics}\ }\textbf {\bibinfo {volume} {113}},\ \bibinfo {pages} {164102} (\bibinfo {year} {2013})}\BibitemShut {NoStop}%
\bibitem [{\citenamefont {Neville}\ \emph {et~al.}(1972)\citenamefont {Neville}, \citenamefont {Hoeneisen},\ and\ \citenamefont {Mead}}]{neville_permittivity_1972}%
  \BibitemOpen
  \bibfield  {author} {\bibinfo {author} {\bibfnamefont {R.~C.}\ \bibnamefont {Neville}}, \bibinfo {author} {\bibfnamefont {B.}~\bibnamefont {Hoeneisen}},\ and\ \bibinfo {author} {\bibfnamefont {C.~A.}\ \bibnamefont {Mead}},\ }\bibfield  {title} {\bibinfo {title} {Permittivity of strontium titanate},\ }\href {https://doi.org/10.1063/1.1661463} {\bibfield  {journal} {\bibinfo  {journal} {Journal of Applied Physics}\ }\textbf {\bibinfo {volume} {43}},\ \bibinfo {pages} {2124} (\bibinfo {year} {1972})}\BibitemShut {NoStop}%
\bibitem [{\citenamefont {Van~Tuan}\ \emph {et~al.}(2019)\citenamefont {Van~Tuan}, \citenamefont {Scharf}, \citenamefont {Wang}, \citenamefont {Shan}, \citenamefont {Mak}, \citenamefont {{\v Z}uti{\'c}},\ and\ \citenamefont {Dery}}]{van_tuan_probing_2019}%
  \BibitemOpen
  \bibfield  {author} {\bibinfo {author} {\bibfnamefont {D.}~\bibnamefont {Van~Tuan}}, \bibinfo {author} {\bibfnamefont {B.}~\bibnamefont {Scharf}}, \bibinfo {author} {\bibfnamefont {Z.}~\bibnamefont {Wang}}, \bibinfo {author} {\bibfnamefont {J.}~\bibnamefont {Shan}}, \bibinfo {author} {\bibfnamefont {K.~F.}\ \bibnamefont {Mak}}, \bibinfo {author} {\bibfnamefont {I.}~\bibnamefont {{\v Z}uti{\'c}}},\ and\ \bibinfo {author} {\bibfnamefont {H.}~\bibnamefont {Dery}},\ }\bibfield  {title} {\bibinfo {title} {Probing many-body interactions in monolayer transition-metal dichalcogenides},\ }\href {https://doi.org/10.1103/PhysRevB.99.085301} {\bibfield  {journal} {\bibinfo  {journal} {Physical Review B}\ }\textbf {\bibinfo {volume} {99}},\ \bibinfo {pages} {085301} (\bibinfo {year} {2019})}\BibitemShut {NoStop}%
\bibitem [{\citenamefont {Liu}\ \emph {et~al.}(2019)\citenamefont {Liu}, \citenamefont {{van Baren}}, \citenamefont {Taniguchi}, \citenamefont {Watanabe}, \citenamefont {Chang},\ and\ \citenamefont {Lui}}]{liu_magnetophotoluminescence_2019}%
  \BibitemOpen
  \bibfield  {author} {\bibinfo {author} {\bibfnamefont {E.}~\bibnamefont {Liu}}, \bibinfo {author} {\bibfnamefont {J.}~\bibnamefont {{van Baren}}}, \bibinfo {author} {\bibfnamefont {T.}~\bibnamefont {Taniguchi}}, \bibinfo {author} {\bibfnamefont {K.}~\bibnamefont {Watanabe}}, \bibinfo {author} {\bibfnamefont {Y.-C.}\ \bibnamefont {Chang}},\ and\ \bibinfo {author} {\bibfnamefont {C.~H.}\ \bibnamefont {Lui}},\ }\bibfield  {title} {\bibinfo {title} {Magnetophotoluminescence of exciton {{Rydberg}} states in monolayer {{WSe$_{2}$}}},\ }\href {https://doi.org/10.1103/PhysRevB.99.205420} {\bibfield  {journal} {\bibinfo  {journal} {Physical Review B}\ }\textbf {\bibinfo {volume} {99}},\ \bibinfo {pages} {205420} (\bibinfo {year} {2019})}\BibitemShut {NoStop}%
\bibitem [{\citenamefont {Liu}\ \emph {et~al.}(2021)\citenamefont {Liu}, \citenamefont {{van Baren}}, \citenamefont {Lu}, \citenamefont {Taniguchi}, \citenamefont {Watanabe}, \citenamefont {Smirnov}, \citenamefont {Chang},\ and\ \citenamefont {Lui}}]{liu_exciton-polaron_2021}%
  \BibitemOpen
  \bibfield  {author} {\bibinfo {author} {\bibfnamefont {E.}~\bibnamefont {Liu}}, \bibinfo {author} {\bibfnamefont {J.}~\bibnamefont {{van Baren}}}, \bibinfo {author} {\bibfnamefont {Z.}~\bibnamefont {Lu}}, \bibinfo {author} {\bibfnamefont {T.}~\bibnamefont {Taniguchi}}, \bibinfo {author} {\bibfnamefont {K.}~\bibnamefont {Watanabe}}, \bibinfo {author} {\bibfnamefont {D.}~\bibnamefont {Smirnov}}, \bibinfo {author} {\bibfnamefont {Y.-C.}\ \bibnamefont {Chang}},\ and\ \bibinfo {author} {\bibfnamefont {C.~H.}\ \bibnamefont {Lui}},\ }\bibfield  {title} {\bibinfo {title} {Exciton-polaron {{Rydberg}} states in monolayer {{MoSe$_{2}$}} and {{WSe$_{2}$}}},\ }\href {https://doi.org/10.1038/s41467-021-26304-w} {\bibfield  {journal} {\bibinfo  {journal} {Nature Communications}\ }\textbf {\bibinfo {volume} {12}},\ \bibinfo {pages} {6131} (\bibinfo {year} {2021})}\BibitemShut {NoStop}%
\bibitem [{\citenamefont {Aslan}\ \emph {et~al.}(2018)\citenamefont {Aslan}, \citenamefont {Deng},\ and\ \citenamefont {Heinz}}]{aslan_strain_2018}%
  \BibitemOpen
  \bibfield  {author} {\bibinfo {author} {\bibfnamefont {B.}~\bibnamefont {Aslan}}, \bibinfo {author} {\bibfnamefont {M.}~\bibnamefont {Deng}},\ and\ \bibinfo {author} {\bibfnamefont {T.~F.}\ \bibnamefont {Heinz}},\ }\bibfield  {title} {\bibinfo {title} {Strain tuning of excitons in monolayer {{WSe$_{2}$}}},\ }\href {https://doi.org/10.1103/PhysRevB.98.115308} {\bibfield  {journal} {\bibinfo  {journal} {Physical Review B}\ }\textbf {\bibinfo {volume} {98}},\ \bibinfo {pages} {115308} (\bibinfo {year} {2018})}\BibitemShut {NoStop}%
\bibitem [{\citenamefont {Schmidt}\ \emph {et~al.}(2016)\citenamefont {Schmidt}, \citenamefont {Niehues}, \citenamefont {Schneider}, \citenamefont {Dr{\"u}ppel}, \citenamefont {Deilmann}, \citenamefont {Rohlfing}, \citenamefont {de~Vasconcellos}, \citenamefont {{Castellanos-Gomez}},\ and\ \citenamefont {Bratschitsch}}]{schmidt_reversible_2016}%
  \BibitemOpen
  \bibfield  {author} {\bibinfo {author} {\bibfnamefont {R.}~\bibnamefont {Schmidt}}, \bibinfo {author} {\bibfnamefont {I.}~\bibnamefont {Niehues}}, \bibinfo {author} {\bibfnamefont {R.}~\bibnamefont {Schneider}}, \bibinfo {author} {\bibfnamefont {M.}~\bibnamefont {Dr{\"u}ppel}}, \bibinfo {author} {\bibfnamefont {T.}~\bibnamefont {Deilmann}}, \bibinfo {author} {\bibfnamefont {M.}~\bibnamefont {Rohlfing}}, \bibinfo {author} {\bibfnamefont {S.~M.}\ \bibnamefont {de~Vasconcellos}}, \bibinfo {author} {\bibfnamefont {A.}~\bibnamefont {{Castellanos-Gomez}}},\ and\ \bibinfo {author} {\bibfnamefont {R.}~\bibnamefont {Bratschitsch}},\ }\bibfield  {title} {\bibinfo {title} {Reversible uniaxial strain tuning in atomically thin {{WSe$_{2}$}}},\ }\href {https://doi.org/10.1088/2053-1583/3/2/021011} {\bibfield  {journal} {\bibinfo  {journal} {2D Materials}\ }\textbf {\bibinfo {volume} {3}},\ \bibinfo {pages} {021011} (\bibinfo {year} {2016})}\BibitemShut {NoStop}%
\bibitem [{\citenamefont {Blundo}\ \emph {et~al.}(2021)\citenamefont {Blundo}, \citenamefont {Yildirim}, \citenamefont {Pettinari},\ and\ \citenamefont {Polimeni}}]{blundo_experimental_2021}%
  \BibitemOpen
  \bibfield  {author} {\bibinfo {author} {\bibfnamefont {E.}~\bibnamefont {Blundo}}, \bibinfo {author} {\bibfnamefont {T.}~\bibnamefont {Yildirim}}, \bibinfo {author} {\bibfnamefont {G.}~\bibnamefont {Pettinari}},\ and\ \bibinfo {author} {\bibfnamefont {A.}~\bibnamefont {Polimeni}},\ }\bibfield  {title} {\bibinfo {title} {Experimental adhesion energy in van der waals crystals and heterostructures from atomically thin bubbles},\ }\href {https://doi.org/10.1103/PhysRevLett.127.046101} {\bibfield  {journal} {\bibinfo  {journal} {Physical Review Letters}\ }\textbf {\bibinfo {volume} {127}},\ \bibinfo {pages} {046101} (\bibinfo {year} {2021})}\BibitemShut {NoStop}%
\bibitem [{\citenamefont {Lyddane}\ \emph {et~al.}(1941)\citenamefont {Lyddane}, \citenamefont {Sachs},\ and\ \citenamefont {Teller}}]{lyddane_polar_1941}%
  \BibitemOpen
  \bibfield  {author} {\bibinfo {author} {\bibfnamefont {R.~H.}\ \bibnamefont {Lyddane}}, \bibinfo {author} {\bibfnamefont {R.~G.}\ \bibnamefont {Sachs}},\ and\ \bibinfo {author} {\bibfnamefont {E.}~\bibnamefont {Teller}},\ }\bibfield  {title} {\bibinfo {title} {On the polar vibrations of alkali halides},\ }\href {https://doi.org/10.1103/PhysRev.59.673} {\bibfield  {journal} {\bibinfo  {journal} {Physical Review}\ }\textbf {\bibinfo {volume} {59}},\ \bibinfo {pages} {673} (\bibinfo {year} {1941})}\BibitemShut {NoStop}%
\bibitem [{\citenamefont {Dyson}(1949)}]{dyson_s_1949}%
  \BibitemOpen
  \bibfield  {author} {\bibinfo {author} {\bibfnamefont {F.~J.}\ \bibnamefont {Dyson}},\ }\bibfield  {title} {\bibinfo {title} {The ${{S}}$ matrix in quantum electrodynamics},\ }\href {https://doi.org/10.1103/PhysRev.75.1736} {\bibfield  {journal} {\bibinfo  {journal} {Physical Review}\ }\textbf {\bibinfo {volume} {75}},\ \bibinfo {pages} {1736} (\bibinfo {year} {1949})}\BibitemShut {NoStop}%
\bibitem [{\citenamefont {Salpeter}\ and\ \citenamefont {Bethe}(1951)}]{salpeter_relativistic_1951}%
  \BibitemOpen
  \bibfield  {author} {\bibinfo {author} {\bibfnamefont {E.~E.}\ \bibnamefont {Salpeter}}\ and\ \bibinfo {author} {\bibfnamefont {H.~A.}\ \bibnamefont {Bethe}},\ }\bibfield  {title} {\bibinfo {title} {A relativistic equation for bound-state problems},\ }\href {https://doi.org/10.1103/PhysRev.84.1232} {\bibfield  {journal} {\bibinfo  {journal} {Physical Review}\ }\textbf {\bibinfo {volume} {84}},\ \bibinfo {pages} {1232} (\bibinfo {year} {1951})}\BibitemShut {NoStop}%
\bibitem [{\citenamefont {Petri{\'c}}\ \emph {et~al.}(2023)\citenamefont {Petri{\'c}}, \citenamefont {Villafa{\~n}e}, \citenamefont {Herrmann}, \citenamefont {Ben~Mhenni}, \citenamefont {Qin}, \citenamefont {Sayyad}, \citenamefont {Shen}, \citenamefont {Tongay}, \citenamefont {M{\"u}ller}, \citenamefont {Soavi}, \citenamefont {Finley},\ and\ \citenamefont {Barbone}}]{petric_nonlinear_2023}%
  \BibitemOpen
  \bibfield  {author} {\bibinfo {author} {\bibfnamefont {M.~M.}\ \bibnamefont {Petri{\'c}}}, \bibinfo {author} {\bibfnamefont {V.}~\bibnamefont {Villafa{\~n}e}}, \bibinfo {author} {\bibfnamefont {P.}~\bibnamefont {Herrmann}}, \bibinfo {author} {\bibfnamefont {A.}~\bibnamefont {Ben~Mhenni}}, \bibinfo {author} {\bibfnamefont {Y.}~\bibnamefont {Qin}}, \bibinfo {author} {\bibfnamefont {Y.}~\bibnamefont {Sayyad}}, \bibinfo {author} {\bibfnamefont {Y.}~\bibnamefont {Shen}}, \bibinfo {author} {\bibfnamefont {S.}~\bibnamefont {Tongay}}, \bibinfo {author} {\bibfnamefont {K.}~\bibnamefont {M{\"u}ller}}, \bibinfo {author} {\bibfnamefont {G.}~\bibnamefont {Soavi}}, \bibinfo {author} {\bibfnamefont {J.~J.}\ \bibnamefont {Finley}},\ and\ \bibinfo {author} {\bibfnamefont {M.}~\bibnamefont {Barbone}},\ }\bibfield  {title} {\bibinfo {title} {Nonlinear ispersion relation and out-of-plane second harmonic generation in {{MoSSe}} and {{WSSe Janus}} monolayers},\ }\href {https://doi.org/10.1002/adom.202300958} {\bibfield
  {journal} {\bibinfo  {journal} {Advanced Optical Materials}\ }\textbf {\bibinfo {volume} {11}},\ \bibinfo {pages} {2300958} (\bibinfo {year} {2023})}\BibitemShut {NoStop}%
\bibitem [{\citenamefont {Wu}\ \emph {et~al.}(2018)\citenamefont {Wu}, \citenamefont {Lovorn}, \citenamefont {Tutuc},\ and\ \citenamefont {MacDonald}}]{wu_hubbard_2018}%
  \BibitemOpen
  \bibfield  {author} {\bibinfo {author} {\bibfnamefont {F.}~\bibnamefont {Wu}}, \bibinfo {author} {\bibfnamefont {T.}~\bibnamefont {Lovorn}}, \bibinfo {author} {\bibfnamefont {E.}~\bibnamefont {Tutuc}},\ and\ \bibinfo {author} {\bibfnamefont {A.~H.}\ \bibnamefont {MacDonald}},\ }\bibfield  {title} {\bibinfo {title} {Hubbard model physics in transition metal dichalcogenide moir\'e bands},\ }\href {https://doi.org/10.1103/PhysRevLett.121.026402} {\bibfield  {journal} {\bibinfo  {journal} {Physical Review Letters}\ }\textbf {\bibinfo {volume} {121}},\ \bibinfo {pages} {026402} (\bibinfo {year} {2018})}\BibitemShut {NoStop}%
\bibitem [{\citenamefont {Pan}\ \emph {et~al.}(2020)\citenamefont {Pan}, \citenamefont {Wu},\ and\ \citenamefont {Das~Sarma}}]{pan_quantum_2020}%
  \BibitemOpen
  \bibfield  {author} {\bibinfo {author} {\bibfnamefont {H.}~\bibnamefont {Pan}}, \bibinfo {author} {\bibfnamefont {F.}~\bibnamefont {Wu}},\ and\ \bibinfo {author} {\bibfnamefont {S.}~\bibnamefont {Das~Sarma}},\ }\bibfield  {title} {\bibinfo {title} {Quantum phase diagram of a moir\'e-{{Hubbard}} model},\ }\href {https://doi.org/10.1103/PhysRevB.102.201104} {\bibfield  {journal} {\bibinfo  {journal} {Physical Review B}\ }\textbf {\bibinfo {volume} {102}},\ \bibinfo {pages} {201104} (\bibinfo {year} {2020})}\BibitemShut {NoStop}%
\bibitem [{\citenamefont {Purdie}\ \emph {et~al.}(2018)\citenamefont {Purdie}, \citenamefont {Pugno}, \citenamefont {Taniguchi}, \citenamefont {Watanabe}, \citenamefont {Ferrari},\ and\ \citenamefont {Lombardo}}]{purdie_cleaning_2018}%
  \BibitemOpen
  \bibfield  {author} {\bibinfo {author} {\bibfnamefont {D.~G.}\ \bibnamefont {Purdie}}, \bibinfo {author} {\bibfnamefont {N.~M.}\ \bibnamefont {Pugno}}, \bibinfo {author} {\bibfnamefont {T.}~\bibnamefont {Taniguchi}}, \bibinfo {author} {\bibfnamefont {K.}~\bibnamefont {Watanabe}}, \bibinfo {author} {\bibfnamefont {A.~C.}\ \bibnamefont {Ferrari}},\ and\ \bibinfo {author} {\bibfnamefont {A.}~\bibnamefont {Lombardo}},\ }\bibfield  {title} {\bibinfo {title} {Cleaning interfaces in layered materials heterostructures},\ }\href {https://doi.org/10.1038/s41467-018-07558-3} {\bibfield  {journal} {\bibinfo  {journal} {Nature Communications}\ }\textbf {\bibinfo {volume} {9}},\ \bibinfo {pages} {5387} (\bibinfo {year} {2018})}\BibitemShut {NoStop}%
\bibitem [{\citenamefont {Wang}\ \emph {et~al.}(2013)\citenamefont {Wang}, \citenamefont {Meric}, \citenamefont {Huang}, \citenamefont {Gao}, \citenamefont {Gao}, \citenamefont {Tran}, \citenamefont {Taniguchi}, \citenamefont {Watanabe}, \citenamefont {Campos}, \citenamefont {Muller}, \citenamefont {Guo}, \citenamefont {Kim}, \citenamefont {Hone}, \citenamefont {Shepard},\ and\ \citenamefont {Dean}}]{wang_one-dimensional_2013}%
  \BibitemOpen
  \bibfield  {author} {\bibinfo {author} {\bibfnamefont {L.}~\bibnamefont {Wang}}, \bibinfo {author} {\bibfnamefont {I.}~\bibnamefont {Meric}}, \bibinfo {author} {\bibfnamefont {P.~Y.}\ \bibnamefont {Huang}}, \bibinfo {author} {\bibfnamefont {Q.}~\bibnamefont {Gao}}, \bibinfo {author} {\bibfnamefont {Y.}~\bibnamefont {Gao}}, \bibinfo {author} {\bibfnamefont {H.}~\bibnamefont {Tran}}, \bibinfo {author} {\bibfnamefont {T.}~\bibnamefont {Taniguchi}}, \bibinfo {author} {\bibfnamefont {K.}~\bibnamefont {Watanabe}}, \bibinfo {author} {\bibfnamefont {L.~M.}\ \bibnamefont {Campos}}, \bibinfo {author} {\bibfnamefont {D.~A.}\ \bibnamefont {Muller}}, \bibinfo {author} {\bibfnamefont {J.}~\bibnamefont {Guo}}, \bibinfo {author} {\bibfnamefont {P.}~\bibnamefont {Kim}}, \bibinfo {author} {\bibfnamefont {J.}~\bibnamefont {Hone}}, \bibinfo {author} {\bibfnamefont {K.~L.}\ \bibnamefont {Shepard}},\ and\ \bibinfo {author} {\bibfnamefont {C.~R.}\ \bibnamefont {Dean}},\ }\bibfield  {title} {\bibinfo {title} {One-dimensional
  electrical contact to a two-dimensional material},\ }\href {https://doi.org/10.1126/science.1244358} {\bibfield  {journal} {\bibinfo  {journal} {Science}\ }\textbf {\bibinfo {volume} {342}},\ \bibinfo {pages} {614} (\bibinfo {year} {2013})}\BibitemShut {NoStop}%
\end{thebibliography}%


\begin{thebibliography}{0}%
\makeatletter
\providecommand \@ifxundefined [1]{%
 \@ifx{#1\undefined}
}%
\providecommand \@ifnum [1]{%
 \ifnum #1\expandafter \@firstoftwo
 \else \expandafter \@secondoftwo
 \fi
}%
\providecommand \@ifx [1]{%
 \ifx #1\expandafter \@firstoftwo
 \else \expandafter \@secondoftwo
 \fi
}%
\providecommand \natexlab [1]{#1}%
\providecommand \enquote  [1]{``#1''}%
\providecommand \bibnamefont  [1]{#1}%
\providecommand \bibfnamefont [1]{#1}%
\providecommand \citenamefont [1]{#1}%
\providecommand \href@noop [0]{\@secondoftwo}%
\providecommand \href [0]{\begingroup \@sanitize@url \@href}%
\providecommand \@href[1]{\@@startlink{#1}\@@href}%
\providecommand \@@href[1]{\endgroup#1\@@endlink}%
\providecommand \@sanitize@url [0]{\catcode `\\12\catcode `\$12\catcode `\&12\catcode `\#12\catcode `\^12\catcode `\_12\catcode `\%12\relax}%
\providecommand \@@startlink[1]{}%
\providecommand \@@endlink[0]{}%
\providecommand \url  [0]{\begingroup\@sanitize@url \@url }%
\providecommand \@url [1]{\endgroup\@href {#1}{\urlprefix }}%
\providecommand \urlprefix  [0]{URL }%
\providecommand \Eprint [0]{\href }%
\providecommand \doibase [0]{https://doi.org/}%
\providecommand \selectlanguage [0]{\@gobble}%
\providecommand \bibinfo  [0]{\@secondoftwo}%
\providecommand \bibfield  [0]{\@secondoftwo}%
\providecommand \translation [1]{[#1]}%
\providecommand \BibitemOpen [0]{}%
\providecommand \bibitemStop [0]{}%
\providecommand \bibitemNoStop [0]{.\EOS\space}%
\providecommand \EOS [0]{\spacefactor3000\relax}%
\providecommand \BibitemShut  [1]{\csname bibitem#1\endcsname}%
\let\auto@bib@innerbib\@empty
\end{thebibliography}%

\section*{\label{sec:acknowledgements}Acknowledgments}
We thank Eduardo Zubizarreta-Casalengua and Ferdinand Menzel for useful discussions. A.B.M acknowledges funding from the International Max Planck Research School for Quantum Science and Technology (IMPRS-QST). M.B. acknowledges funding from the A. von Humboldt Foundation. We gratefully acknowledge funding from the Deutsche Forschungsgemeinschaft (DFG, German Research Foundation) via Germany’s Excellence Strategy (MCQST, EXC-2111/390814868, and e-conversion, EXC-2089/1-390776260). J.J.F. also acknowledges the BMBF for funding via projects 16K15Q027, 13N15760, 13N16214, as well as the DFG via INST 95/1719-1, FI 947/6-1, INST 95/1496-1, FI 947/5-1, FI 947/8-1, DI 2013/5-1 and DI 2013/5-2. K.M. also acknowledges the DFG via the project PQET (INST 95/1654-1). Furthermore, we acknowledge the Bavarian Science Ministry for funding via the Munich Quantum Valley, Nequs, and IQ-Sense projects.
Work at the University of Rochester was supported by the Department of Energy, Basic Energy Sciences, Division of Materials Sciences and Engineering under Award No. DE-SC0014349.
S.T. acknowledges primary support from DOE-SC0020653 (materials synthesis), NSF CMMI 1825594 (NMR and TEM studies), NSF ECCS 2052527 (electrical testing), DMR 2111812 (optical testing), and CMMI 2129412 (scaling).
K.W. and T.T. acknowledge support from the JSPS KAKENHI (Grant Numbers 20H00354 and 23H02052) and World Premier International Research Center Initiative (WPI), MEXT, Japan.

\section*{\label{sec:contributions}Author contributions}
A.B.M. and M.B. conceived and managed the research. A.B.M., and L.G. fabricated the devices. A.B.M., L.G., M.M.P., and M.B. performed the optical measurements. A.B.M. and M.B. analyzed the results. J.J.F. and K.M. obtained third party funding and provided experimental and nanofabrication infrastructure. D.V.T. and H.D. developed the models and performed the calculations. M.E. and S.T. grew WSe$_{2}$, WS$_{2}$, and MoSe$_{2}$ bulk crystals. K.W. and T.T. grew bulk hBN crystals. All authors discussed the results and contributed to the writing of the paper.\newline

\section*{\label{sec:interests}Competing interests}
The authors declare no competing interests.

\section*{\label{sec:si}Supplementary information}
Supplementary Figs. 1–6 and Theoretical Methods.

\end{document}


\title{Supplementary information:\\ Breakdown of the static dielectric screening approximation of Coulomb interactions in atomically thin semiconductors}

\author{Amine Ben Mhenni}
\email{amine.ben-mhenni@wsi.tum.de}
\affiliation{%
    Walter Schottky Institute, TUM School of Natural Sciences, and MCQST, Technical University of Munich, Munich, Germany
}%

\author{Dinh Van Tuan}%

\affiliation{%
    Department of Electrical and Computer Engineering, University of Rochester, Rochester, NY, United States.
}%

\author{Leonard Geilen}%
\affiliation{%
    Walter Schottky Institute, TUM School of Natural Sciences, and MCQST, Technical University of Munich, Munich, Germany
}%

\author{Marko M. Petrić}%
\affiliation{%
    Walter Schottky Institute, TUM School of Computation, Information and Technology, and MCQST, Technical University of Munich, Munich, Germany
}%

\author{Melike Erdi}%
\affiliation{%
    School for Engineering of Matter, Transport and Energy, Arizona State University, Tempe, AZ, United States.
}%

\author{Kenji Watanabe}%
\affiliation{%
 Research Center for Electronic and Optical Materials, National Institute for Materials Science, 1-1 Namiki, Tsukuba 305-0044, Japan
}%

\author{Takashi Taniguchi}%
\affiliation{%
 Research Center for Materials Nanoarchitectonics, National Institute for Materials Science,  1-1 Namiki, Tsukuba 305-0044, Japan
}%

\author{Sefaattin Tongay}%
\affiliation{%
 School for Engineering of Matter, Transport and Energy, Arizona State University, Tempe, AZ, United States.
}%

\author{Kai Müller}%
\affiliation{%
    Walter Schottky Institute, TUM School of Computation, Information and Technology, and MCQST, Technical University of Munich, Munich, Germany
}%

\author{Nathan P. Wilson}%
\affiliation{%
    Walter Schottky Institute, TUM School of Natural Sciences, and MCQST, Technical University of Munich, Munich, Germany
}%

\author{Jonathan J. Finley}%
\email{finley@wsi.tum.de}
\affiliation{%
    Walter Schottky Institute, TUM School of Natural Sciences, and MCQST, Technical University of Munich, Munich, Germany
}%

\author{Hanan Dery}%
\affiliation{%
    Department of Electrical and Computer Engineering, University of Rochester, Rochester, NY, United States.
}%
\affiliation{%
    Department of Physics and Astronomy, University of Rochester, Rochester, NY, United States.
}%

\author{Matteo Barbone}%
\email{matteo.barbone@wsi.tum.de}
\affiliation{%
    Walter Schottky Institute, TUM School of Computation, Information and Technology, and MCQST, Technical University of Munich, Munich, Germany
}%

\date{\today}

\maketitle

\tableofcontents
\clearpage

\section{Blueshift of X$^{0}$ in MoSe$_{2}$ and WS$_{2}$\label{sec:s4}}

\begin{figure*}[hb]
\includegraphics[width=1\textwidth]{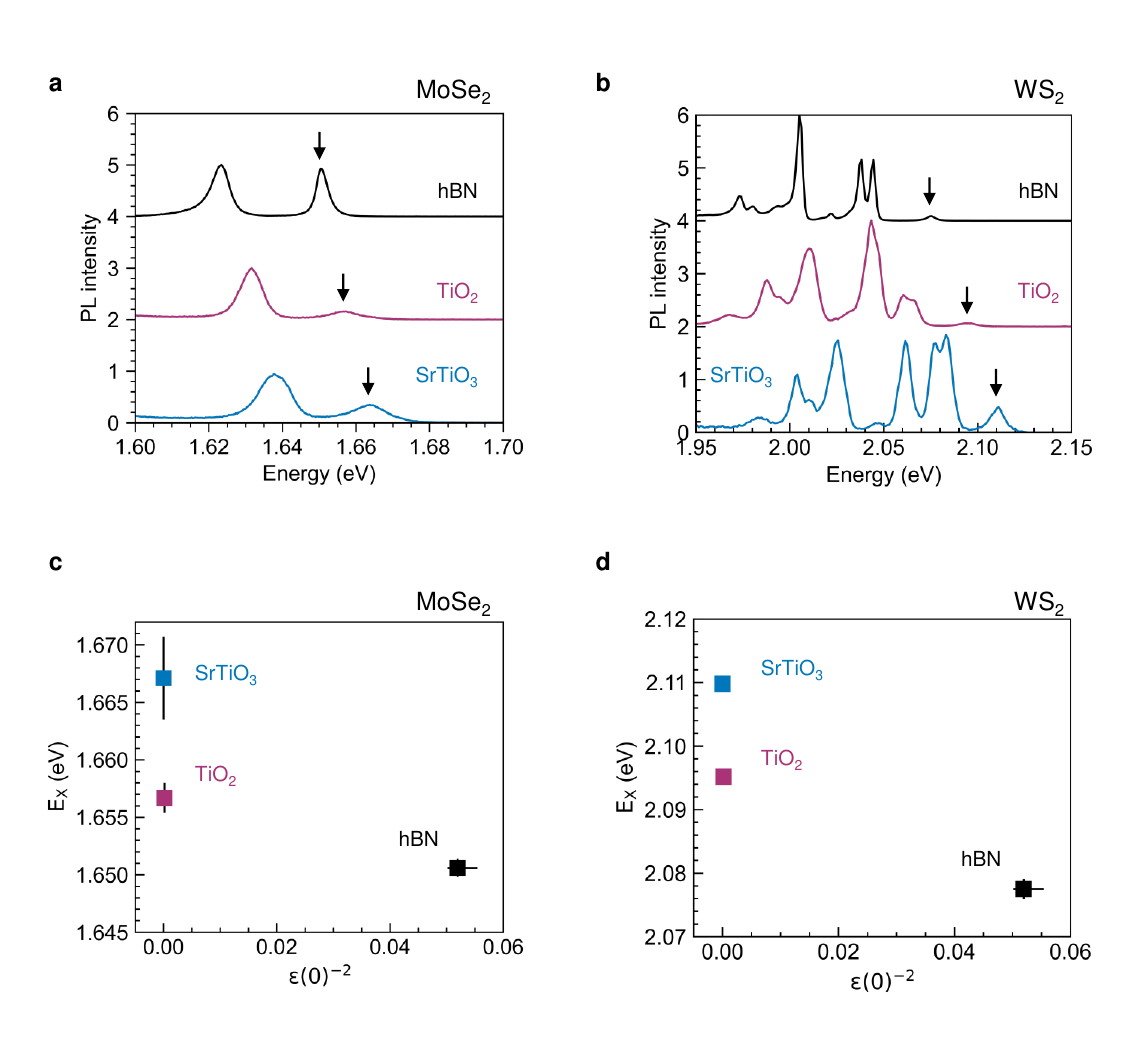}
\caption{\label{sifig:analysis}\textbf{Effect of the dielectric screening on X$^{0}$ in MoSe$_{2}$ and WS$_{2}$.}
\newline
\textbf{a}, PL spectra of ungated monolayer MoSe$_{2}$ in the hBN, TiO$_{2}$, and SrTiO$_{3}$ dielectric configurations. \textbf{b}, PL spectra of ungated monolayer WS$_{2}$ in the hBN, TiO$_{2}$, and SrTiO$_{3}$ dielectric configurations. The spectral position of X$^{0}$ is indicated by an arrow in both figures. Both for MoSe$_{2}$ and WS$_{2}$, X$^{0}$ blueshifts with a higher static dielectric constant. Here, the shift resulting from charge doping was not accounted for. However, the magnitude of the blueshift is much larger than what can be explained by just charge doping effects. \textbf{c} and \textbf{d}, $\mathrm{E_{X}}$ as a function of $\mathrm{\varepsilon(0)^{-2}}$ for MoSe$_{2}$  (\textbf{c}) and WS$_{2}$ (\textbf{d}). The mean value of $\mathrm{E_{X}}$ across large areas on more than one sample was taken for each data point. The standard deviation is plotted.
}
\end{figure*}

\clearpage

\section{Optical data analysis\label{sec:s0}}

\begin{figure*}[hb]
\includegraphics[width=1\textwidth]{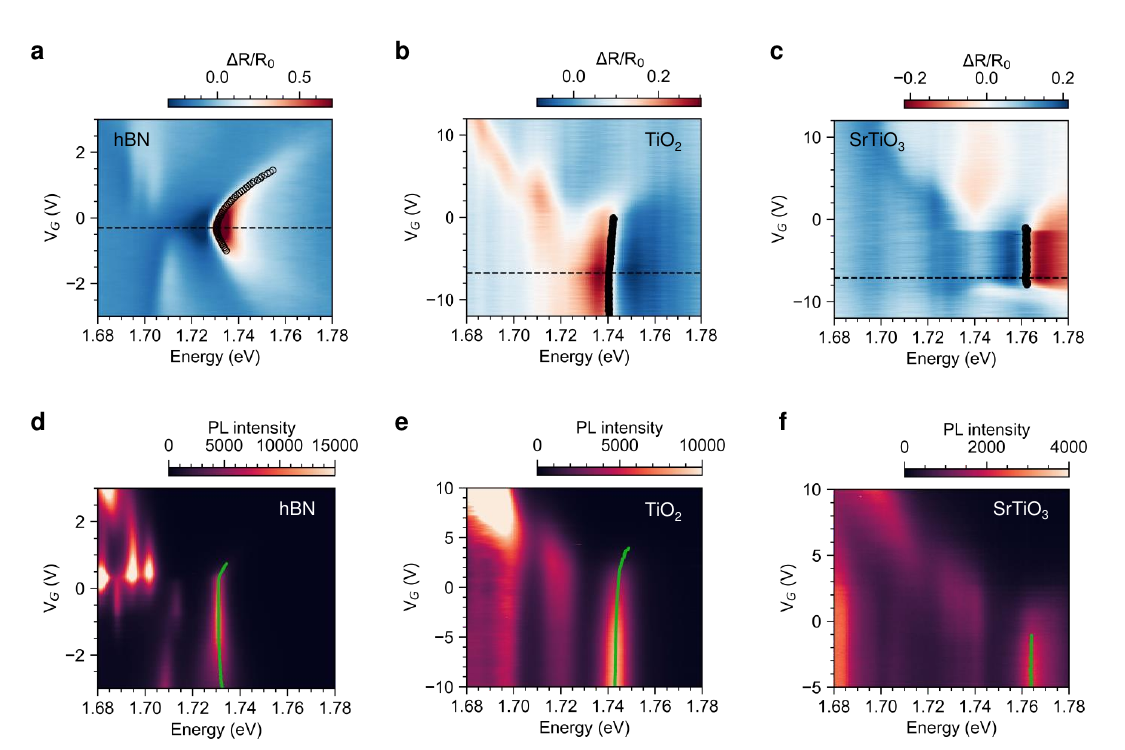}
\caption{\label{sifig:analysis}\textbf{Reflection contrast and PL data analysis.}
\newline
Gate-dependent reflection contrast spectra from the hBN (\textbf{a}), TiO$_{2}$ (\textbf{b}), and SrTiO$_{3}$ (\textbf{c}) device. A dispersive Lorentzian \cite{smolenski_signatures_2021} was used to fit X$^{0}$ in the vicinity of the charge-neutral region. The energy of the dispersive Lorentzian fit is overlaid (in black). The charge neutrality point is extracted via the minimum of $\mathrm{E_{X}}$ and is indicated by a horizontal dashed line. Gate-dependent PL spectra from the hBN (\textbf{d}), TiO$_{2}$ (\textbf{e}), and SrTiO$_{3}$ (\textbf{f}). A Lorentzian was used to fit X$^{0}$. The energy of the Lorentzian fit is overlaid (in green).
}
\end{figure*}

\clearpage

\section{Blueshift of X$^{0}$ due to charge doping\label{sec:s2}}

\begin{figure*}[hb]
\includegraphics[width=1\textwidth]{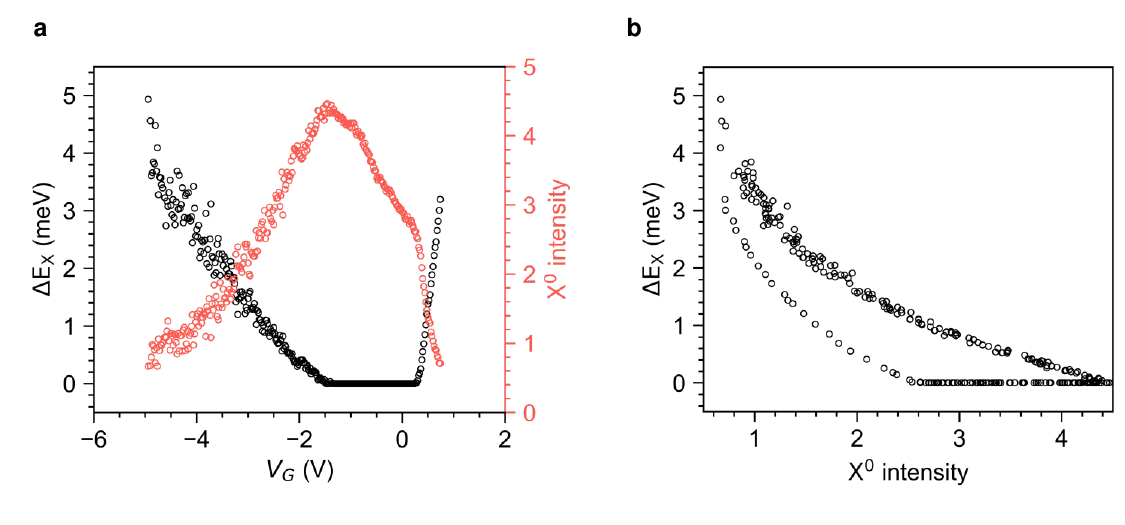}
\caption{\textbf{Effect of charge doping on $\mathrm{E_{X}}$ for monolayer WSe$_{2}$ in the hBN configuration. }
\newline
\textbf{a}, $\mathrm{E_{X}}$ change from its value at charge neutrality (black), and X$^{0}$ intensity (coral) as a function of the gate voltage (V$_{G}$). The data is extracted from the fit discussed in Supplimentary Fig. \ref{sifig:analysis}. \textbf{b}, $\mathrm{E_{X}}$ change from its value at charge neutrality as a function of its intensity. The data shows that at higher charge doping, X$^{0}$ loses almost an order of magnitude of its maximum intensity when it has already blueshifted by $5$ meV, consistent with Ref. \cite{van_tuan_probing_2019}.}
\end{figure*}

\clearpage

\section{Rydberg series of WSe$_{2}$ in the hBN device\label{sec:s1}}

\begin{figure*}[hb]
\includegraphics[width=1\textwidth]{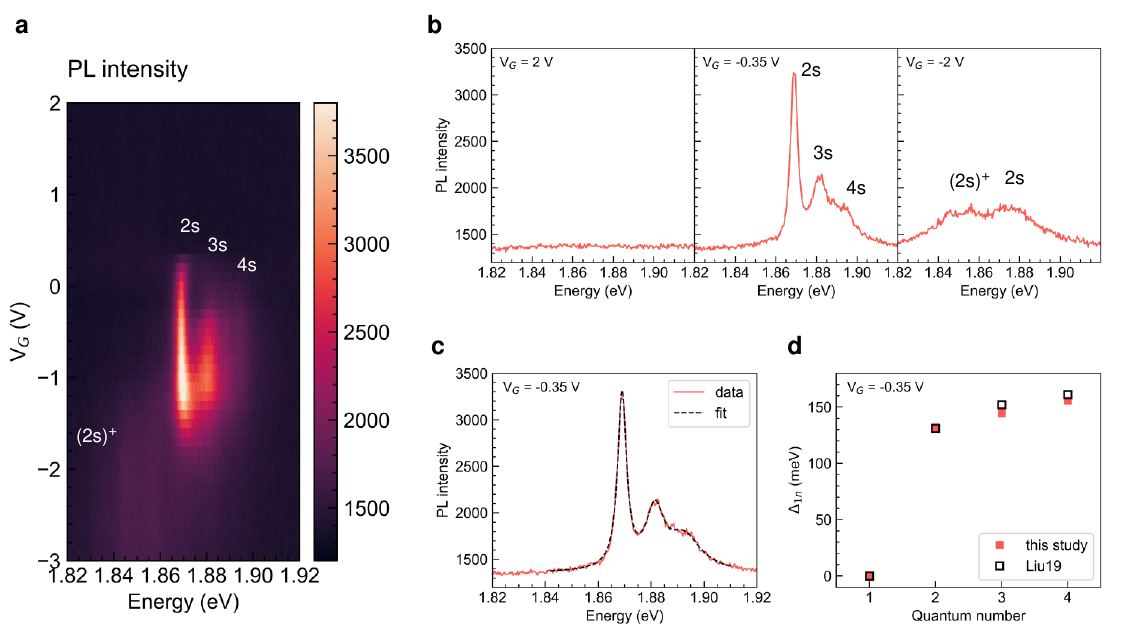}
\caption{\textbf{PL of the exciton Rydberg series of monolayer WSe$_{2}$ in the hBN configuration. }
\newline
\textbf{a}, Gate-dependent PL spectra of the hBN device showing a well-resolved exciton Rydberg series of WSe$_{2}$, namely the $2s$, $3s$, and $4s$ excitons, in addition to the positively charged $2s$ resonance $(2s)^{+}$. \textbf{b}, PL spectra extracted from (\textbf{a}) in three different charge-doping regimes: electron doping ($2$ V), charge neutrality (-$0.35$ V), and hole doping ($-2$ V). \textbf{c}, Fitting of the Rydberg series using $3$ Lorentzians. \textbf{d}, Energy difference of the n$^{th}$ Rydberg resonance and the $1s$ resonance showing excellent agreement with Ref. \cite{liu_magnetophotoluminescence_2019}, reflecting high sample quality.}
\end{figure*}

\clearpage

\section{Neutral exciton energy distribution \label{sec:s3}}

\begin{figure*}[hb]
\includegraphics[width=1\textwidth]{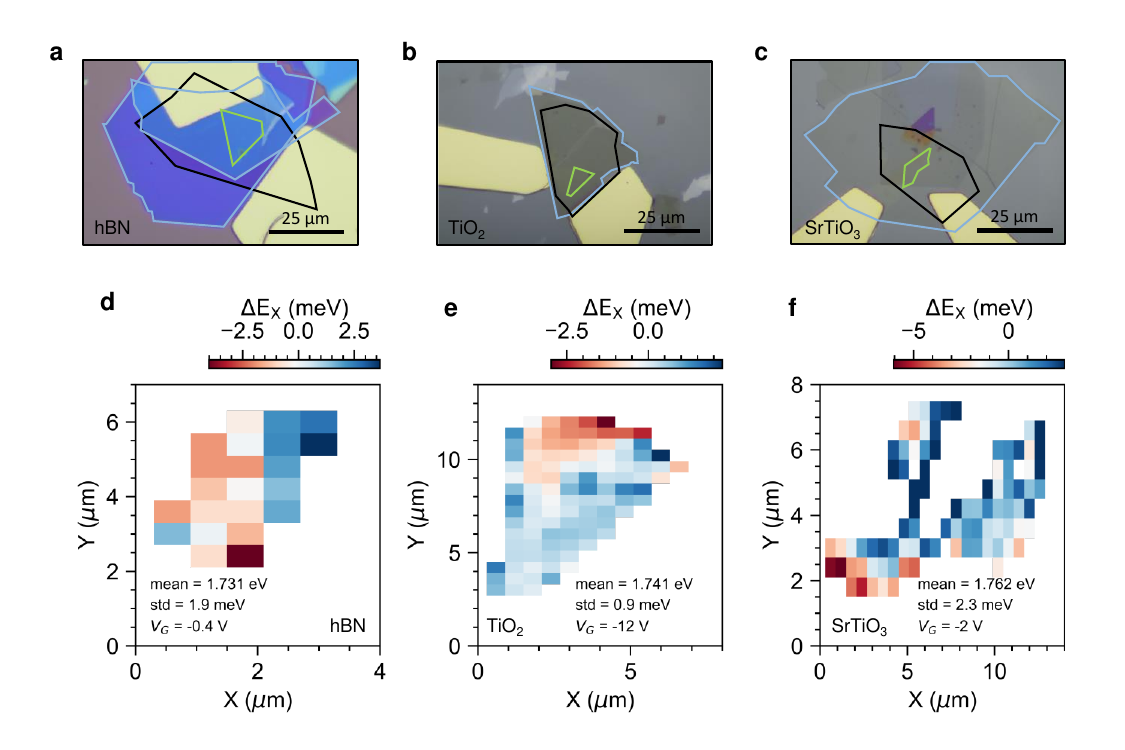}
\caption{\textbf{Neutral exciton energy distribution from PL sample maps.}
\newline
(\textbf{a-c}), Optical micrographs of the hBN (\textbf{a}), TiO$_{2}$ (\textbf{b}), and SrTiO$_{3}$ (\textbf{c}) devices. (\textbf{d-f}), Two-dimensional PL maps of the $\mathrm{E_{X}}$ variation from the mean energy in the hBN (\textbf{d}), TiO$_{2}$ (\textbf{e}), and SrTiO$_{3}$ (\textbf{f}) devices. A threshold on X$^0$ PL intensity was set to filter out pixels containing flake edges and electrically unconnected regions of the monolayers. The data shows an $\mathrm{E_{X}}$ standard deviation lower than $3$ meV in all devices, an order of magnitude lower than the observed blueshift.}
\end{figure*}

\clearpage

\section{Temperature-dependence of the SrTiO$_{3}$ device and saturation of the dielectric effect \label{sec:s5}}
\begin{figure*}[hb]
\includegraphics[width=1\textwidth]{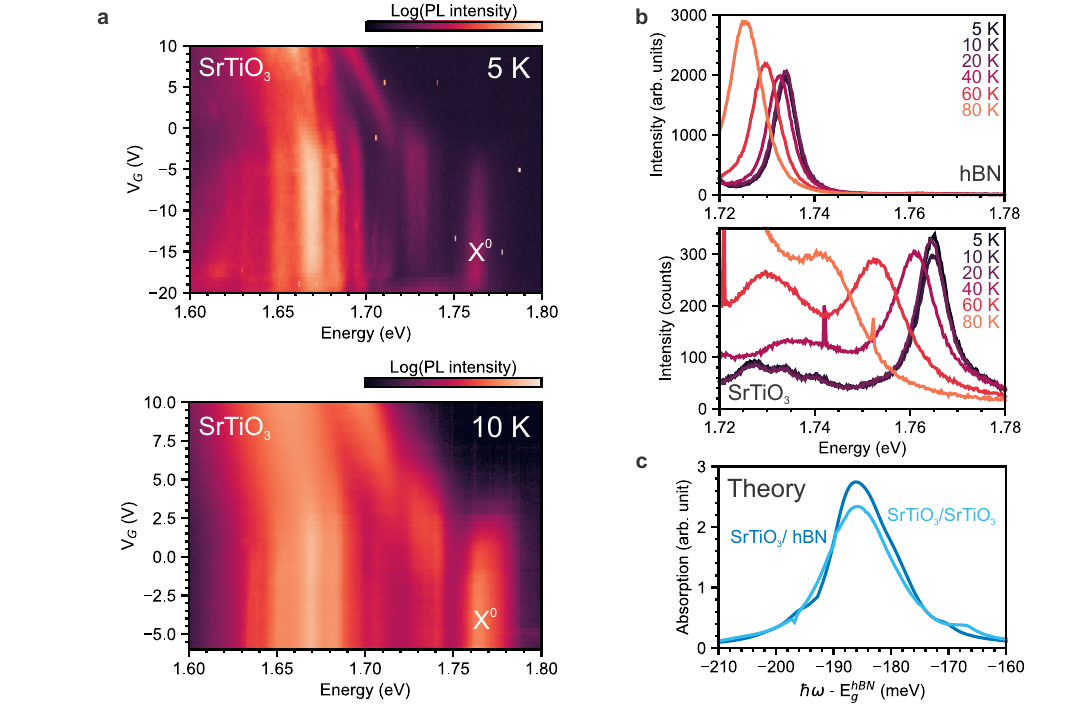}
\caption{\textbf{Temperature-dependence of the SrTiO$_{3}$ device and saturation of the dielectric effect.}
\newline
(\textbf{a}), Gate-dependent PL spectra of monolayer WSe$_{2}$ in the SrTiO$_{3}$ configuration for two different temperatures: $5$ K (top panel) and $10$ K (bottom panel). 
Although, the dielectric constant of SrTiO$_{3}$ decreases rapidly going from lower temperature to higher temperature \cite{Neville1972}, X$^{0}$ energy doesn't show a temperature-dependent shift within the experimental uncertainty going from $5$ K to $10$ K. 
This demonstrates the saturation of the dielectric effect. 
(\textbf{b}), Temperature-dependent PL spectra of X$^{0}$ of the hBN (top panel) and the SrTiO$_{3}$ device (bottom panel) taken close to charge neutrality at: $5$ K, $10$ K, $20$ K, $40$ K, $60$ K, and $60$ K.
In the hBN configuration, X$^{0}$ exhibits an almost constant energy up to $40$ K, and it shifts by about $9$ meV at $80$ K relative to $5$ K. 
This effect stems from the decrease of the band gap at higher temperatures and is consistent with previous reports \cite{Nagler2018}.
In the SrTiO$_{3}$ configuration, X$^{0}$ exhibits a constant energy up to $20$ K, although the dielectric constant of SrTiO$_{3}$ decreases rapidly as the temperature increases \cite{Neville1972}. This corroborates the saturation of the dielectric effect: starting from a certain value (the dielectric constant of SrTiO$_{3}$ at $20$ K), any further increase in the dielectric constant doesn't induce an X$^{0}$ shift.
At higher temperatures, X$^{0}$ exhibits a more significant shift compared to the hBN device, namely: $4$ meV, $12$ meV, and more than $20$ meV at $40$ K, $60$ K, and $80$ K, respectively.
In addition to the known shift stemming from the band gap reduction due to thermal effects, a further mechanism is probably at play. 
Starting from $20$K, the decrease in the dielectric constant of SrTiO$_{3}$ when the temperature is increases results in an effective decrease in the screening and thus a redshift of X$^{0}$.
This presents a further corroboration of the blueshifting of X$^{0}$ for stronger-screening environements and offers a new route for the tunability of the optoelectronic properties of atomically thin semiconducting materials with applications in nanophotonics \cite{Weber2023}.
(\textbf{c}), Absorption spectra of the SrTiO$_{3}$/hBN device in addition to a hypothetical SrTiO$_{3}$/SrTiO$_{3}$ configuration. 
The latter spectrum doesn’t exhibit any distinguishable shift relative to the SrTiO$_{3}$/hBN configuration (SrTiO3 as a bottom and hBN as a top dielectric), indicating that the latter configuration already constitutes a limiting case of the dynamical screening.
\newline
Note: The gate-dependent spectra at $10$ K (\textbf{a}, bottom panel) is the same shown in the main text.}
\end{figure*}

\clearpage

\section{Theoretical methods\label{sec:s6}}

\subsection{Theory}
The static Coulomb interaction between charge particles in the monolayer is obtained by solving the Poisson Equation with the appropriate structure geometry. The simulated structure geometry is a monolayer with thickness $d$ sandwiched between top and bottom layers with dielectric constants $\epsilon_\text{t}$ and $\epsilon_\text{b}$.  
The TMD monolayer is modeled as three atomic sheets with polarizabilities $ \chi_+$ for the central Tungsten (W) sheet and $ \chi_-$  for the top and bottom Selenium (Se)  ones, displaced by $\pm d/4$ from the center. The model was developed in Ref. \cite{VanTuan_PRB18} and has been employed to study several problems \cite{VanTuan_PRL22,VanTuan_PRB22,VanTuan_PRL19}. The resulting static potential for the interaction between two charges in the monolayer is
\begin{equation}
V(q) = \frac{2 \pi e^2}{A \,\,  \epsilon(q) \,\, q},
\label{Eq:Static}
\end{equation}
where $A$ is the area of the system, $q$ is the transferred crystal momentum during the interaction, and the static permittivity function of the structure is given by
\begin{equation}\label{Eq:DiFv2}
\epsilon(q)=\frac{1}{2}\left[\frac{N_t(q)}{D_t(q)}+\frac{N_b(q)}{D_b(q)}\right].
\end{equation}
Defining $p_j \equiv (\epsilon_j-1)/(\epsilon_j+1)$ for the top and bottom dielectric constants ($j=b/t$), we get that 
\begin{eqnarray}\label{Eq:DiFv2def}
D_j(q) &=& 1+q\ell_- -q\ell_- (1+p_j)\text{e}^{-\frac{qd}{2}} - (1-q \ell_- ) p_j \text{e}^{-qd}, \nonumber \\
N_j(q) &=& \left(1+q\ell_-\right)\left(1+q\ell_+\right)  + \left[\left(1-p_j\right)-\left(1+p_j\right)q\ell_+\right]q\ell_-\text{e}^{-\frac{qd}{2}} + (1-q\ell_-)(1-q\ell_+ )p_j\text{e}^{-qd}.
\end{eqnarray}
where $\ell_\pm = 2 \pi  \chi_\pm$. 

\vspace{4mm}

The Coulomb potential in Eq.~(\ref{Eq:Static}) becomes frequency dependent, $V(q) \rightarrow V(q,\omega)$, by using dynamical polarization parameters  in Eq.~(\ref{Eq:DiFv2def}). Namely,
\begin{eqnarray}
p_{j} \rightarrow p_{j}(\omega) = \frac{ \epsilon_{j}(\omega)-1}{\epsilon_{j}(\omega)+1}\,\,,\qquad \text{where}\,\,\,\,\,\,\,\,\,\,\epsilon_j(\omega) = \epsilon_{\infty} \,\,\prod_i \frac{\omega_{i,j,\text{LO}}^2 - \omega^2}{\omega_{i,j,\text{TO}}^2 - \omega^2}\,\,. 
\label{eq:LST}
\end{eqnarray}
$\epsilon_{j}(\omega)$ is the frequency-dependent permittivity of the $j=b/t$ layer, as given by Eq.~(1) of the main manuscript. The ratio between the static and high-frequency permittivities is the celebrated Lyddane–Sachs–Teller relation $\epsilon_j(\omega=0)/\epsilon_j(\omega=\infty) \equiv  \epsilon_{j,0}/\epsilon_{j,\infty} =  \prod_i \omega_{i,j,\text{LO}}^2 /\omega_{i,j,\text{TO}}^2 $. The index $i$ runs over the optical-phonon modes of the $j=b/t$ layer, where $\omega_{i,j,\text{LO/TO}}$ is the associated frequency of the longitudinal/transverse optical lattice vibration.  

\vspace{4mm}

The dynamical Coulomb potential has singularities at the phonon frequencies. To circumvent this difficulty when solving the Bethe-Salpeter Equation (BSE) or evaluating the self-energies, we use finite-temperature Green's function formalism in which real frequencies are replaced by imaginary and discrete Matsubara frequencies \cite{MahanBook}. Namely, $V({\bf q}, \omega)$  is replaced with $V({\bf q}, z-z')$, where $z$ and $z'$ are imaginary Matsubara energies of fermions before and after the interaction. Their discretized energy form is $(2\ell+1)\pi i k_\text{B} T$, where $\ell$ is an integer and T is temperature. Consequently, the positive real number $\omega^2$ in Eq.~(\ref{eq:LST}) is replaced by $(z-z')^2$ which is a negative real number. Rather than having singularities, the permittivity function $\epsilon_j(z-z')$  is now monotonously decaying from $\epsilon_0$ to $\epsilon_\infty$ as $z-z'$ departs from 0.   

\subsection{Dynamical Self-Energy}

The self-energy of an electron in  the conduction ($c$) or valence band (v) is calculated  from a self-consistent solution of the Dyson Equation
\begin{equation}
\Sigma_i({\bf k},z ) = - \frac{1}{\beta} \sum_{{\bf q},z'} G_i({\bf k- q}, z' ) V({\bf q}, z - z'). 
\label{eq:DynaSelfE}
\end{equation}
where  $\beta^{-1}=k_BT$, $i=\{\text{c,v}\}$, and the Green's function is
\begin{eqnarray}
G_i({\bf k},z ) &= &\frac{G_i^0({\bf k},z )}{1 - G_i^0({\bf k},z ) \Sigma_i({\bf k},z )}   = \frac{1}{z - \varepsilon_i({\bf k}) + \mu- \Sigma_i({\bf k},z ) }.  
\end{eqnarray}
${\bf k}$ and $ \mu$ are the electron momentum and its chemical potential, respectively.   $ \varepsilon_\text{c}({\bf k}) =  E_\text{g} + {\hbar^2 k^2}/{2 m_c} $ is the energy dispersion of the electron in the conduction band and  $ \varepsilon_\text{v}({\bf k}) =   {\hbar^2 k^2}/{2  m_\text{v}} $  is the corresponding one in the valence band. 

One difficulty of self-energy calculations is the divergence of the sum over $\bf q$. We illustrate this point by using the non-dynamical bandgap renormalization (BGR), wherein the potential becomes $V(\mathbf{q},z-z') \rightarrow V(\mathbf{q})$ and the sum over Matsubara energies in Eq.~(5)  is rendered straightforward, $\Sigma_i(\mathbf{k},z) = \Sigma_i(\mathbf{k}) = \pm \tfrac{1}{2}\sum_{\mathbf{q}}V(\mathbf{q})$. The $\pm$ denotes the self-energy of an electron (hole) in the conduction (valence) band.  The 2D potential $V(q) $ scales as $q^{-2}$ in the short wavelength limit, resulting in a logarithmic divergence of the sum over $\bf q$.    We circumvent this problem by choosing a reference TMD structure with respect to which energy shifts are calculated.  Without loss of generality, we choose a reference system whose corresponding potential  $V_0(q)$ is evaluated with the static permittivity function, as given by Eqs.~(\ref{Eq:Static})-(\ref{Eq:DiFv2def}),  using $\epsilon_\text{t}  = \epsilon_\text{b} = 3.8$. The BGR of a given system with respect to the reference system is then given by
\begin{eqnarray}
\!\!\!\!\!\!\!\!\!\!\!\!\tilde{\Sigma}_i({\bf k},z ) &=& - \frac{1}{\beta} \sum_{{\bf q},z'} G_i({\bf k+ q}, z' )  \times \left[ V({\bf q}, z - z') -V_0({\bf q}) \right], \,\,\,\,\, \,\,\,\, \,\,\,\
\label{eq:DynaSelfE1}
\end{eqnarray} 
where the free electron Green's function now becomes
\begin{eqnarray}
G_i({\bf k},z )= \frac{1}{z - \varepsilon_i({\bf k}) + \mu- \tilde{\Sigma}_i({\bf k},z ) }.
\label{eq:GreenE} 
\end{eqnarray}
The dynamical self-energy $\tilde{\Sigma}_i({\bf k},z )$ can be self-consistently  calculated  from Eqs.~(\ref{eq:DynaSelfE1}) and (\ref{eq:GreenE})  using an iterative method. 

\subsection{ Dynamical Bethe-Salpeter Equation }

The BSE is an equation for bound states between two particles. Its  dynamical version is used here to describe the  interaction between electron and hole excited by light with negligible momentum \cite{Scharf_JPCM19,Haug_SchmittRink_PQE84,VanTuan_PRX17}
  
\begin{eqnarray}  
 G({\bf k},z,\Omega)= G^0({\bf k},z,\Omega) \,\,\, + \,\,\,\frac{1}{\beta}\sum_{{\bf  q},z'}G^0({\bf k},z,\Omega) V({\bf q},z-z')G({\bf k+q},z',\Omega), 
 \label{Bethe-Salpeter1}
\end{eqnarray}
where the Green's function of a free electron-hole pair is given by 
\begin{eqnarray}
 G^0({\bf k},z,\Omega) &=&  \frac{1}{\Omega - z - \varepsilon_\text{c}({\bf k})  - \tilde{\Sigma}_\text{c}({\bf k},\Omega - z )   + \mu }  \times  \frac{1}{ z + \varepsilon_\text{v}({\bf k})+  \tilde{\Sigma}_\text{v}({\bf k}, z ) - \mu} . 
 \label{Eq:FreeGreen}
\end{eqnarray} 
$\Omega$,  an even (bosonic) imaginary Matsubara energy, is related to the energy of the photon exciting the electron-hole pair. Equation~(\ref{Bethe-Salpeter1}) can be solved using the iterative method, the same method as the one used for calculating the self-energies in   Eqs.~(\ref{eq:DynaSelfE1})-(\ref{eq:GreenE}). One can notice that solutions of different bosonic frequencies $\Omega$ are decoupled, and therefore, equations of different $\Omega$s can be solved independently. The solutions are  then used to find the contracted pair function 
\begin{equation}
g({\bf k},\Omega) = -\beta^{-1} \sum_{z} G({\bf k},z,\Omega). 
\label{Eq:BSECon}
\end{equation}
The final step is to analytically continue the contracted pair function to the real-frequency axis, $g({\bf k},\Omega \rightarrow \omega + i \delta)$, using the Pad\'{e} approximation technique  \cite{VanTuan_PRX17,Scharf_JPCM19,Vidberg_JLTP1977}.     The real-frequency pair function is related to optical absorption by 
\begin{equation}
\text{A}(\omega) = - \sum_{{\bf k}} \text{Im} \left[ g({\bf k},\Omega \rightarrow \omega + i \delta)\right],
\label{Eq:Ana1}
\end{equation}
where $\delta$ is broadening parameter which might include effects  of finite exciton lifetime, scattering off  impurities, and thermal fluctuations. 
Note that temperature in this formalism sets the energy resolution of Matsubara frequencies and is not related to the broadening of resonance peaks which is controlled by $\delta$. In this work, we keep $\delta = 3$~meV for the sake of simplicity.

In the non-dynamical regime (static permittivity), the potential and  self-energies are frequency independent.  The BSE in Eq.~(\ref{Bethe-Salpeter1}) can be further contracted, yielding 
\begin{equation}
 \!\!\! g({\bf k},\Omega)= g^0({\bf k},\Omega)- \sum_{{\bf  q}}g^0({\bf k},\Omega)\,\,  V({\bf q})\,\, g({\bf k+q},\Omega), 
 \label{Eq:StaBSE}
\end{equation} 
and the corresponding function of a free electron-hole pair is given by \cite{MahanBook,Haug_SchmittRink_PQE84}
\begin{eqnarray}
g^0({\bf k},\Omega) &=& -\beta^{-1} \sum_{z} G^0({\bf k},z,\Omega) = \frac{ f_{\text{v}}({\bf k}) - f_{\text{c}}({\bf k})}{\Omega    + \varepsilon_\text{v}({\bf k}) +  \tilde{\Sigma}_\text{v}({\bf k}) - \varepsilon_\text{c}({\bf k}) - \tilde{\Sigma}_\text{c}({\bf k})  }. \,\,\,\,\,\,\,\,\,\, \,\,\,\, \label{Eq:Free}
\end{eqnarray}
$f_{\text{c(v)}}({\bf k})$ is the Fermi-Dirac distribution function for electrons in the conduction (valence) band. 

\subsection{Distinction between static and dynamical calculations}
When dealing with self-energies of free electrons or holes (i.e., when they are not part of an excitonic complex), their calculated self-energies are very close whether they are calculated with frequency-dependent permittivity or with static permittivity using the low-frequency dielectric constant $\epsilon_0$. Namely, free charge particles in the monolayer are also screened by phonons in the polar dielectric materials. 

Unlike the case of free charge particles, the self-energy of the electron in the exciton is associated with $(\Omega - z)$ whereas that of the hole with $z$, as shown by Eq.~(\ref{Eq:FreeGreen}). The bosonic frequency $\Omega$ is related to the photon energy, which is of the order of the bandgap energy, a large value compared with phonon or binding energies. Consequently, the self-energy of at least one of the exciton's components asymptotically approaches the value of the self-energy when calculated non-dynamically with $\epsilon_\infty$. An alternative view is that the energy difference between the exciton components is encoded as a time-dependent phase factor $\exp(iE_g t/\hbar)$, which leads to a dominant contribution from the high-frequency part of the dielectric function to the exciton's BGR. 

\subsection{Dynamical effects in the optical spectrum}

The dynamical potential  $V({\bf q}, z-z')$ affects both the BGR and binding energy. To decouple these effects, we focus first on the binding energy by neglecting the self-energy terms in the BSE (i.e., $\tilde{\Sigma}_\text{c/v}({\bf k},z ) = 0$ in Eq.~(\ref{Bethe-Salpeter1})).  The resulting absorption spectra of hBN-WSe$_2$-SrTiO$_{3}$ structures are shown in Fig.~3c of the main text. For comparison, we have also calculated the absorption spectra with static permittivities using low-frequency dielectric constants (dashed lines)  and high-frequency ones (dash-dotted lines). The exciton binding energy is $E_\text{b} = 122 $~meV when using the dynamical potential in hBN-WSe$_2$-SrTiO$_{3}$. Corresponding values of the non-dynamical calculations are $E_\text{b}^0 =104$~meV and $E_\text{b}^\infty = 181$~meV.  The dynamical binding energy is closer to the one calculated with $\epsilon_0$,  meaning that screening of the interaction between the electron and hole is dominated by the low-frequency part of the dielectric function. 

\vspace{4mm}

Next, we include the self-energy in the calculations of the absorption spectra and consider the competition between BGR and binding energy. The energy blueshift can be recovered if dynamical effects are considered through the potential $V({\bf q}, z-z')$, self-energies $\tilde{\Sigma}_\text{c}({\bf k},\Omega- z )$ and $\tilde{\Sigma}_\text{v}({\bf k},z )$, and if replacing the encapsulating materials involve a large change in $\epsilon_0$ and a small change in $\epsilon_\infty$. Figure~3b of the main text shows the resulting absorption spectra of three different structures:  hBN-WSe$_2$-hBN,  hBN-WSe$_2$-TiO$_2$, and  hBN-WSe$_2$-SrTiO$_{3}$. In agreement with the experimental results, replacing the supporting hBN layer with TiO$_2$ leads to energy blueshift, which is further increased when SrTiO$_{3}$ is used as support. The opposite energy-shift trends of calculations with static and dynamical permittivities can be explained as follows. The binding energy is mainly dominated by the low-frequency part of the dielectric function, where the change is from $\epsilon^\text{hBN}_0 = 4.9$ to $ \epsilon^{\text{TiO}_2}_0 \sim 200$ and then to $\epsilon^\text{SrTiO$_{3}$}_0= 25000$. As a result, the change in binding energy is relatively significant. On the other hand, the self-energies of the electron and hole in the exciton have larger contribution from the high-frequency part, where the change is from $\epsilon^\text{hBN}_\infty = 3.8$ to $\epsilon^\text{TiO$_2$}_\infty \sim \epsilon^\text{SrTiO$_{3}$}_\infty \sim 6$. As a result, the BGR effect is relatively mitigated. The confluence of both trends is that the energy redshift from BGR is smaller than the energy blueshift from binding energy ($  \Delta  E_\text{g} < |\Delta E_\text{b}|$), leading to overall energy blueshift of the exciton resonance.  

\clearpage

